\documentclass[aps, pra, reprint, superscriptaddress, nofootinbib]{revtex4-2}

\usepackage{amsmath, amssymb, amsfonts}
\usepackage{graphicx}
\usepackage[colorlinks=true, linkcolor=blue, citecolor=blue, urlcolor=blue]{hyperref}
\usepackage{xcolor}
\usepackage{braket}
\usepackage{multirow}
\usepackage{booktabs}

\usepackage{verbatim}

\begin{document}

\title{Performance analysis of cat and GKP breeding protocols \\ with squeezed single-photon input}

\author{Yoonha Kim}
\affiliation{Department of Physics, University of Seoul, Seoul 02504, Republic of Korea}
\affiliation{Center for Quantum Technology, Korea Institute of Science and Technology (KIST), Seoul 02792, Republic of Korea}

\author{Jaehak Lee}
\affiliation{Center for Quantum Technology, Korea Institute of Science and Technology (KIST), Seoul 02792, Republic of Korea}
\affiliation{Division of Quantum Information, KIST School, Korea University of Science and Technology (UST), Seoul 02792, Republic of Korea}

\begin{abstract}
Homodyne breeding is a promising optical route for generating large cat states and Gottesman--Kitaev--Preskill (GKP) grid states, using readily available squeezed-cat states as inputs. We analyze the performance of protocols, using squeezed single-photon seeds, $\hat{S}(r)\ket{1}$, as squeezed-cat resources and propagating them through the same cat-to-GKP breeding pipeline. For a symmetric tree with $N=2^{m_1}$ seeds, the heralded resource state reduces to a state with a polynomial-Gaussian wavefunction controlled by a single effective width, achieving near-unit fidelity with the desired cat resource state. Our analysis yields closed-form even-cat fidelities and an inverse-design rule for the seed squeezing. We apply these resources to GKP breeding, and find a large overlap with a practical GKP state. We also investigate the effects of practical imperfections, including finite homodyne acceptance windows and photon loss. This study shows that squeezed single-photon inputs provide a viable resource assignment for homodyne cat and GKP breeding, replacing the demanding requirement for high-resolution photon-number-resolving heralding with an on-demand single-photon source.
\end{abstract}

\maketitle

\section{Introduction}
Practical quantum computation requires physical errors to be suppressed at the logical level through quantum error correction and fault-tolerant architectures~\cite{shor_scheme_1995, steane_error_1996, preskill_reliable_1998, fowler_surface_2012}. Bosonic systems are well suited to this task, since a single harmonic-oscillator mode offers a large Hilbert space in which quantum information can be redundantly encoded~\cite{gottesman_encoding_2001, terhal_towards_2020}. Optical modes are a particularly attractive bosonic platform: photons are natural carriers for transmission, can be processed at high bandwidth, and are compatible with integrated photonic circuits~\cite{kok_linear_2007, slussarenko_photonic_2019, wang_integrated_2020, aghaee_rad_scaling_2025}. Realizing this promise, however, hinges on preparing the required nonclassical resource states with sufficient quality and rate.

Continuous-variable (CV) optical quantum information is encoded in the quadrature degrees of freedom, where squeezing, linear optics, displacements, and homodyne detection form a high-precision experimental toolbox~\cite{braunstein_quantum_2005, weedbrook_gaussian_2012, furusawa_quantum_2011}. This Gaussian toolbox is not sufficient on its own: Gaussian operations cannot distill Gaussian entanglement, and universal CV computation requires at least one non-Gaussian ingredient~\cite{lloyd_quantum_1999, bartlett_efficient_2002, eisert_distilling_2002, niset_no-go_2009, mari_positive_2012}. The generation of optical non-Gaussian states is therefore a decisive bottleneck, as available optical nonlinearities are not sufficiently strong to produce such states directly.

Two central non-Gaussian resources are Gottesman--Kitaev--Preskill (GKP) grid states and the Schr\"{o}dinger cat states. Cat states are important non-Gaussian resources for optical quantum information processing, offering several useful applications~\cite{Glancy:08, lund_conditional_2004, ourjoumtsev_generation_2007, zhang_quantifying_2021}. At the same time, they serve as an intermediate resource for generating GKP grid states~\cite{vasconcelos_all-optical_2010, weigand_generating_2018, konno_logical_2024}. GKP states encode a qubit in a phase space lattice and correct small displacement errors through stabilizer measurements~\cite{gottesman_encoding_2001, glancy_error_2006, menicucci_fault-tolerant_2014, fukui_high-threshold_2018}. Their appeal for optical quantum computing lies in their compatibility with measurement-based CV architectures: once available, GKP qubits embedded in CV cluster states enable deterministic gate operations as well as the implementation of fault-tolerant computation~\cite{menicucci_fault-tolerant_2014, fukui_high-threshold_2018, bourassa_blueprint_2021, larsen_deterministic_2021, aghaee_rad_scaling_2025}. GKP states have been realized in trapped ions and superconducting circuits~\cite{fluhmann_encoding_2019, campagne-ibarcq_quantum_2020}, and have recently been demonstrated for propagating light~\cite{konno_logical_2024, larsen_integrated_2025}, though reaching the state quality required for fault tolerance in optical modes remains especially demanding.

A prominent optical route to these states is breeding: small non-Gaussian inputs are combined by Gaussian operations and conditioned on homodyne outcomes, amplifying seed cat states into larger cats~\cite{lund_conditional_2004, laghaout_amplification_2013, etesse_experimental_2015, sychev2017enlargement, simon_experimental_2024} or into finite grid states~\cite{vasconcelos_all-optical_2010, weigand_generating_2018, konno_logical_2024}. This approach is attractive since, given the input cats, the breeding stages require only Gaussian operations and homodyne conditioning.

In optical schemes, non-Gaussianity is usually supplied by photon-counting measurement. Photon subtraction heralds a small cat state by detecting a photon tapped from squeezed light~\cite{ourjoumtsev_generation_2006, neergaard-nielsen_generation_2006, takahashi_generation_2008, gerrits_generation_2010}. Its generalization to photon-number-resolving (PNR) heralding enhances the success probability and attainable amplitude~\cite{takase_generation_2021, anteneh_machine_2024}. Further, PNR-based methods have also been applied directly to optical GKP-state synthesis~\cite{takase_gottesman-kitaev-preskill_2023, hanamura_beyond_2026}.
These photon-counting schemes are effective for generating non-Gaussian states, but they rely on nondeterministic measurements and impose demanding requirements on photon-number resolution and detection efficiency.

Meanwhile, a single photon is already a non-Gaussian state, while squeezing provides an adjustable Gaussian degree of freedom before the breeding stages. Using single photons as cat precursors is itself long established~\cite{ourjoumtsev_generation_2007,simon_experimental_2024}; what is less clear is how such squeezed single-photon seeds perform as the input layer of a complete cat-to-GKP pipeline.
Recent developments in deterministic quantum-dot sources providing high-quality, clocked single photons~\cite{ding_high-efficiency_2025, maring_versatile_2024, rickert_high_2025} motivate starting from an on-demand single-photon resource rather than a probabilistically heralded cat precursor.

In this work, we employ a squeezed single-photon state, $\hat{S}(r)\ket{1}$, as the elementary seed of homodyne cat and GKP breeding and analyze its performance throughout the full breeding pipeline.
For the symmetric tree with $N=2^{m_1}$ inputs, the output reduces to a polynomial-Gaussian family governed by a single effective width, giving closed-form even-cat fidelities and an inverse-design rule for the seed squeezing. By incorporating these resources into GKP breeding, we place the seed, the bred cat, and the resulting GKP state within a unified quantitative framework. We also make an analysis considering practical imperfections, including finite homodyne windows without sampling and photon loss. By supplying the non-Gaussianity directly at the source level, this approach reduces the reliance on demanding photon-counting hardware and the switching requirements of iterative heralding, while remaining compatible with clocked single-photon sources and fast homodyne readout.

The paper is organized as follows. Section~\ref{breeding_and_resource} develops the analytic description of homodyne cat breeding from squeezed single-photon inputs and applies the resulting bred resources to GKP generation. Section~\ref{finite_window} extends the treatment to finite homodyne acceptance windows through an analytic outcome polynomial description, and Sec.~\ref{loss_tolerance} examines the tolerance to photon loss. Section~\ref{comparison} compares the resulting resource assignment with PNR-based schemes, and Section~\ref{conclusion} concludes.

\section{Breeding protocol and resource analysis}
\label{breeding_and_resource}
In this section, we develop an analytic description of homodyne breeding from squeezed single-photon seeds. We first establish, at the seed level, the regime in which a squeezed single photon provides a high-fidelity approximation to a small-amplitude squeezed cat state, identifying the amplitude beyond which seed-level preparation alone is insufficient. We then formulate the cat-breeding protocol in the quadrature wavefunction representation and show that, for the symmetric circuit, the heralded output reduces to a polynomial--Gaussian family parametrized by a single effective width $\sigma_{\mathrm{eff}}$. 

\begin{figure*}
\centering
\includegraphics[width=0.99\textwidth]{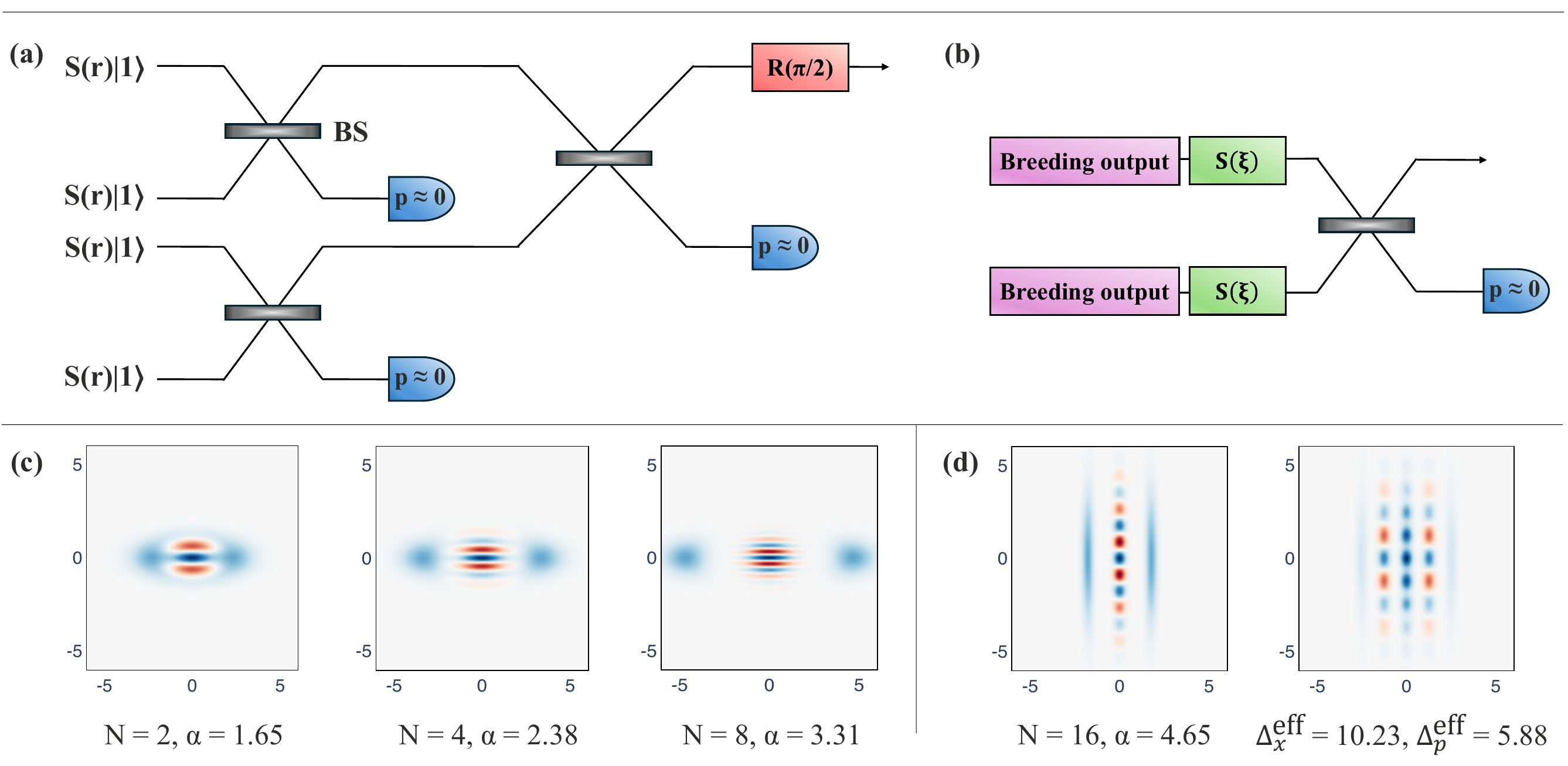}
\caption{
Overview of the resource-generation scheme. 
(a) Four-seed homodyne cat-breeding circuit. Seed states are mixed via 50:50 beamsplitters (BSs), followed by homodyne conditioning measurements. The circuit can be generalized to $m_1$ breeding layers using $N=2^{m_1}$ seeds.
(b) Generic breeding step used in the GKP generation stage. GKP breeding can be repeated over $m_2$ layers, where the input resources are bred cat states for the first layer and GKP states generated in the preceding layer for subsequent layers. Squeezing is required for lattice matching but can be incorporated into the seed-state preparation.
(c) Wigner functions of the bred cat resources for $N=2,4,8$.
(d) Wigner functions of the spacing-matched $N=16,\quad \zeta=1.31$ squeezed-cat resource and the resulting finite-energy GKP ($d_x=d_p=\sqrt{2\pi}$) state, with effective squeezings in dB.
}
\label{fig1_overview}
\end{figure*}

The overall breeding schemes are described in Fig.~\ref{fig1_overview}. We consider the cat breeding with $N=2^{m_1}$ identical seeds, where the circuit is analytically tractable using a single parameter characterizing the effective width. In particular, the even-cat fidelity can be obtained in closed form, and its optimization determines the seed squeezing used in the subsequent stage. We then feed these analytically chosen resources into the GKP-breeding stage after applying an additional squeezing $\zeta$ to match the target lattice spacing, and evaluate the resulting states.

\subsection{Seed approximation}
The squeezed single-photon states used as seed states are given by
\begin{equation}
    \hat{S}(r)\ket{1} = \exp\left[\frac{r}{2}(\hat{a}^2-\hat{a}^{\dagger2})\right]\ket{1},
\end{equation}
where $\ket{1}$ is the single-photon state, $\hat{S}(r)$ is the squeezing operation with the squeezing parameter $r$, and $\hat{a}$ and $\hat{a}^\dagger$ are the annihilation and creation operators, respectively. Since odd Schr\"{o}dinger cat states are known to approach single-photon states in the small-amplitude ($\alpha \ll 1$) limit~\cite{Glancy:08, lund_conditional_2004}, squeezed single-photon states may replace squeezed cat state seeds that are conventionally prepared probabilistically. Therefore, we perform a seed-level analysis to quantify the fidelity and parameter regime in which this replacement is justified.

Taking $\alpha\in\mathbb{R}$ and $\alpha>0$, Schr\"{o}dinger cat states are defined as 
\begin{equation}
\begin{aligned}
    &\ket{\mathrm{Cat}_\pm(\alpha)} = \frac{1}{\mathcal{N}_\pm}(\ket{\alpha} \pm \ket{-\alpha}) , \\
    &\mathcal{N}_\pm = \sqrt{2[1\pm\exp(-2\alpha^2)]}.
\end{aligned}
\end{equation}
Here, the signs $+$ and $-$ denote, respectively, the even and odd cat states. Further, squeezed cat states are given by $\ket{\mathrm{SCS}_\pm(\alpha, \zeta)} = \hat{S}(\zeta)\ket{\mathrm{Cat}_\pm(\alpha)}$. Since the seed and target squeezing operations are applied along the same quadrature, the seed overlap depends only on the relative squeezing parameter $s=r-\zeta$ such that $\hat{S}^\dagger(\zeta)\hat{S}(r) = \hat{S}(s)$. The fidelity of the seed state to the odd squeezed cat state is then given by
\begin{equation}
\begin{aligned}
    F_\mathrm{seed}(s, \alpha) &= \left|\bra{\mathrm{Cat}_-(\alpha)}\hat{S}^\dagger(\zeta)\hat{S}(r)\ket{1}\right|^2 \\
    &= \left|\bra{\mathrm{Cat}_-(\alpha)}\hat{S}(s)\ket{1}\right|^2.
\end{aligned}
\end{equation}
The seed overlap can be evaluated in closed form~\cite{Glancy:08}; in the present convention it reads
\begin{equation}
    F_{\mathrm{seed}}(s,\alpha)=\frac{2\alpha^2}{\cosh^3 s(1-e^{-2\alpha^2})}e^{-\alpha^2(1+\tanh s)}.
\end{equation}

Because the input and target squeezings collapse into the single relative squeezing parameter $s$, the seed level admits a one-dimensional operating-point analysis: for each target amplitude $\alpha$ there is an optimal $s^*(\alpha)$ at which $F_\mathrm{seed}$ is maximized. Optimizing over $s$ gives $F_\mathrm{seed} \geq 0.99$ up to $\alpha \simeq 1.2$, with $s^* \simeq -0.426$ at the upper edge of this window. The amplitude beyond this point cannot be achieved with high fidelity using the seed state alone; hence an amplification process is required, which can be realized through the breeding protocol.

\subsection{$\mathbf{N=2}$ Cat breeding}

The first step in the cat breeding protocol is to mix $S(r_1)\ket{1}\otimes S(r_2)\ket{1}$ on a beam splitter parametrized by the angle $\theta$, followed by the homodyne projection of mode~2 onto the $p$ quadrature. We first consider the $p=0$ outcome and later generalize the analysis to a finite acceptance window. A final rotation by $\pi/2$ is applied to the heralded output state to express it in the same quadrature frame as the target cat state $\ket{\mathrm{Cat}_+(\alpha)}$. To obtain the conditional output state analytically, we describe the protocol in the $\hat{x}$-quadrature wavefunction representation. The quadrature operators are related to the annihilation operator by $\hat{a}=(\hat{x}+i\hat{p})/\sqrt{2}$ where $[\hat{x},\hat{p}]=i$ and $\hbar=1$. The squeezing operator introduced above then satisfies $\hat{S}^\dagger(r)\hat{x}\hat{S}(r)=\hat{x}e^{-r}$ and $\hat{S}^\dagger(r)\hat{p}\hat{S}(r)=\hat{p}e^{r}$, so that $r>0$ corresponds to squeezing along the $\hat{x}$-quadrature. The $\hat{x}$-quadrature wavefunction of the squeezed single-photon state is then given by
\begin{equation}
    \psi_{r}(x)=\bra{x}\hat{S}(r)\ket{1} = \sqrt{2}\pi^{-\frac{1}{4}} e^\frac{3r}{2} x \exp\left(-\frac{1}{2}e^{2r}x^2\right),
\end{equation}
where $\ket{x}$ is the eigenstate of $\hat{x}$.

After the breeding protocol, including $\pi/2$ rotation, the wavefunction of the resulting state is given by (see Appendix \ref{AppendixA1} for details)
\begin{equation}
    \psi^{(2)}_{\mathrm{out}}(x) = \frac{2}{\sqrt{3}\pi^{1/4}\sigma_{\mathrm{eff}}^{5/4}} x^2 \exp\left(-\frac{x^2}{2\sigma_{\mathrm{eff}}}\right),
    \label{N2_output}
\end{equation}
where the effective width is determined by
\begin{equation}
    \sigma_{\mathrm{eff}}=\frac{e^{2(r_1+r_2)}}{e^{2r_1}\sin^2\theta+e^{2r_2}\cos^2\theta}.
\end{equation}
States of this form are known to closely approximate squeezed cat states: Ref.~\cite{ourjoumtsev_generation_2007} showed that mixing a photon-number state $\ket{n}$ with vacuum and conditioning on a homodyne outcome yields a wavefunction of the form $x^n e^{-x^2/2}$, which well approximates a squeezed cat state. The breeding protocol considered here produces the same polynomial-Gaussian family from a different route, in which squeezed single-photon seeds are combined through successive beam-splitter and homodyne-conditioning steps rather than prepared as a single high Fock state. Unlike the protocol in Ref.~\cite{ourjoumtsev_generation_2007}, where the Gaussian width is fixed, the width here is tunable, and the bred resource therefore admits an analytic treatment: the fidelity to the cat state can be obtained in closed form as a function of both $\sigma_{\mathrm{eff}}$ and the target amplitude $\alpha$.

Since the output wavefunction $\psi^{(2)}_{\mathrm{out}}(x)$ is an even function of $x$, its overlap with an odd cat state vanishes. This parity structure follows directly from the homodyne-conditioning step: the projection onto the outcome $p=0$ acts as an even-parity filter on the measured mode. The two squeezed single-photon seeds carry an even total photon-number parity, which is preserved by the beam splitter, so the heralded output has even parity. Thus, the benchmark for the two-seed output is its fidelity to the even cat state $\ket{\mathrm{Cat}_+(\alpha)}$, which we define as
\begin{equation}
    F^{(2)}_+(\sigma_{\mathrm{eff}},\alpha)=\left|\braket{\mathrm{Cat}_+(\alpha)|\psi^{(2)}_{\mathrm{out}}}\right|^2.
\end{equation}
Evaluating the overlap using the normalized wavefunction in Eq.~\eqref{N2_output} yields a closed form 
\begin{equation}
\begin{aligned}
    F^{(2)}_+(\sigma_{\mathrm{eff}}, \alpha)=&\frac{16\sqrt{\sigma_{\mathrm{eff}}}(2\sigma_{\mathrm{eff}}\alpha^2+\sigma_{\mathrm{eff}}+1)^2}{3(\sigma_{\mathrm{eff}}+1)^5}\\
    &\times \frac{e^{2\sigma_{\mathrm{eff}}\alpha^2/(\sigma_{\mathrm{eff}}+1)}}{e^{2\alpha^2}+1}.
    \label{N2_Fidelity}
\end{aligned}
\end{equation}
Equation~\eqref{N2_Fidelity} shows that the circuit parameters $r_1$, $r_2$, and $\theta$ enter the normalized output only through the single effective width $\sigma_{\mathrm{eff}}$; the remaining freedom in $(r_1,r_2,\theta)$ specifies only a circuit realization of the same state. Maximizing $F^{(2)}_+(\sigma_{\mathrm{eff}}, \alpha)$ over $\sigma_{\mathrm{eff}}$ for each target amplitude $\alpha$ shows that the fidelity remains above $0.99$ up to $\alpha^{(2)}_{\mathrm{max}}\simeq1.65$, with the threshold optimum occurring at $\sigma^*_{\mathrm{eff}}\simeq2.38$. In Fig.~\ref{fig2_landscape}(a), we plot the fidelity landscape, assuming $ r_1 = r_2 \equiv r $ and a balanced beam splitter. The region with fidelity above 0.99 is clearly enclosed by the red contour. 

The two-seed circuit is the elementary building block of the breeding protocol. Applying the same beam splitter transformation and homodyne conditioning on $p=0$ to a pair of such states raises the polynomial order once more, and iterating this step extends the scheme to $N=2^{m_1}$ squeezed single-photon seeds. We formulate this recursion in the pre-rotation representation, as detailed in the following subsection.

\subsection{$\mathbf{N=2^{m_1}}$ Cat breeding}
We now generalize the two-seed analysis to $N=2^{m_1}$ seeds. For compact notation, we refer to the output after $m_1$ cat-breeding layers, corresponding to $N=2^{m_1}$ squeezed single-photon seeds, as an $m_1$-cat. Following the all-optical breeding architecture of Vasconcelos~\textit{et al.}~\cite{vasconcelos_all-optical_2010}, the successive breeding rounds are organized as a balanced binary tree. By the same parity argument, every breeding round remains in the even-parity sector, so the benchmark remains the fidelity to $\ket{\mathrm{Cat}_+(\alpha)}$.

In the recursive construction, the unrotated conditional wavefunction is used as the input to the next breeding round. The $\pi/2$ phase-space rotation is therefore applied only once, to the final output, so that all intermediate states are kept in the same representation. In addition, we restrict our analysis to the symmetric configuration, in which all seed squeezing parameters are equal and all beam splitters are balanced $(\theta=\pi/4)$. The Gaussian part then reproduces itself after each round: if the two inputs have width $\sigma_{\mathrm{eff}}$, the heralded output has the same width. The recursion therefore acts only on the polynomial factor, while the Gaussian envelope can be kept fixed throughout.
The final heralded state is then (see Appendix \ref{AppendixA2} for details)
\begin{equation}
\begin{aligned}
    &\psi^{(N)}_{\mathrm{out}}(x)=\mathcal{N}^{(N)}_{\sigma_{\mathrm{eff}}}x^N\exp\left(-\frac{x^2}{2\sigma_{\mathrm{eff}}}\right),\\
    &\mathcal{N}^{(N)}_{\sigma_{\mathrm{eff}}}=\frac{2^{N/2}}{\sqrt{(2N-1)!!}\pi^{1/4}\sigma_{\mathrm{eff}}^{(2N+1)/4}}.
    \label{N_output}
\end{aligned}
\end{equation}
This expression extends the two-seed output in Eq.~\eqref{N2_output} to $N=2^{m_1}$ seeds, retaining the polynomial-Gaussian form.

For the even-parity output in Eq.~\eqref{N_output}, the fidelity to $\ket{\mathrm{Cat}_+(\alpha)}$, denoted by $F^{(N)}_+(\sigma_{\mathrm{eff}},\alpha)$, can be evaluated analytically, with the explicit expression derived in Appendix \ref{AppendixA3}.
With the resource size $N$ and target amplitude $\alpha$ specified, the optimization is only over $\sigma_{\mathrm{eff}}$. The outcome of this one-dimensional optimization is summarized by
\begin{equation}
    \bar{F}^{(N)}_+(\alpha)=\max_{\sigma_{\mathrm{eff}}}F^{(N)}_+(\sigma_{\mathrm{eff}},\alpha),
\end{equation}
from which we define the largest amplitude supported by the $\bar{F}^{(N)}_+(\alpha)\ge0.99$ region as
\begin{equation}
    \alpha^{(N)}_{\mathrm{max}}=\max\{\alpha:\bar{F}^{(N)}_+(\alpha)\ge0.99\}.
\end{equation}

\begin{figure}
\includegraphics[width=\linewidth]{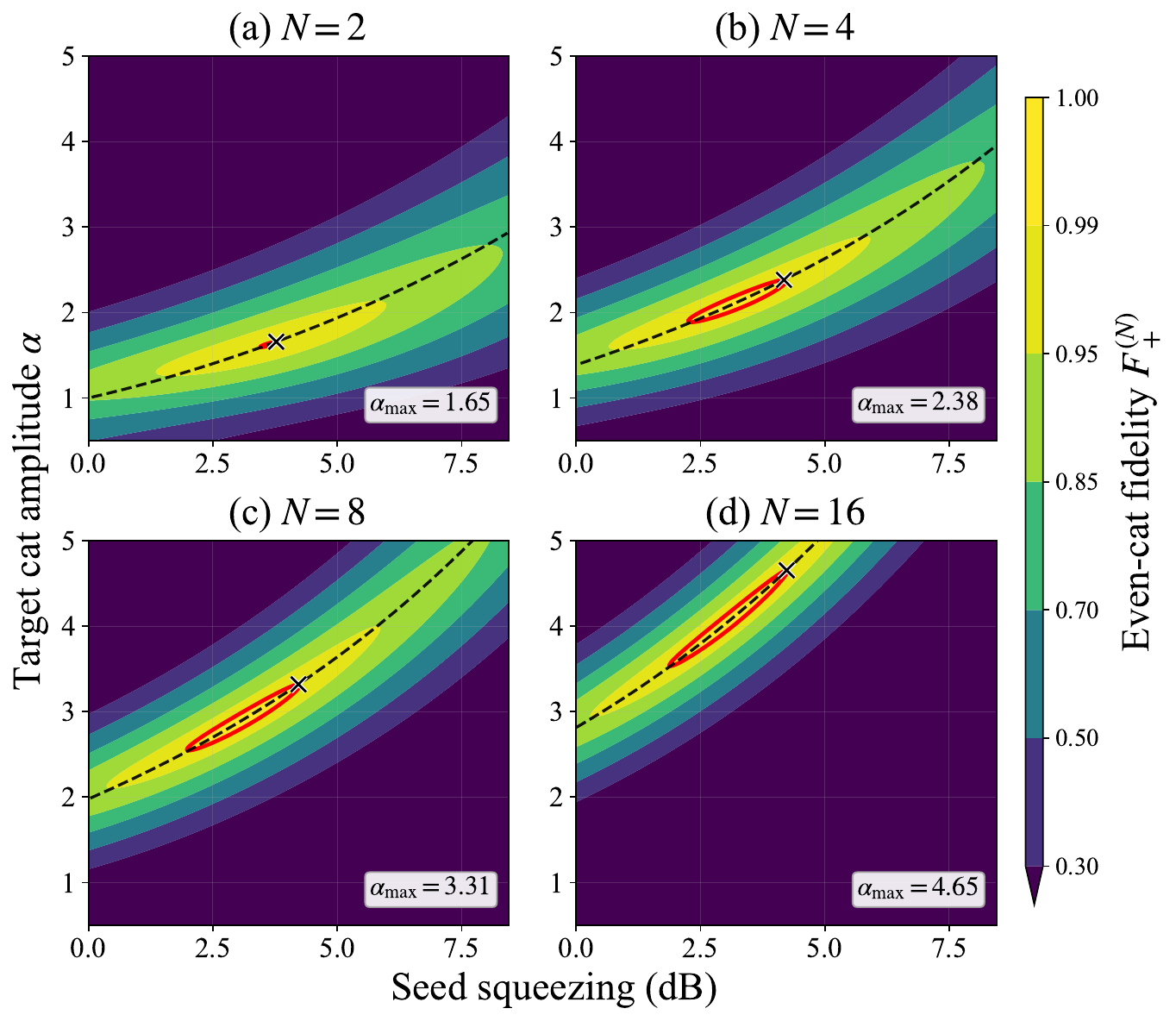}
\caption{Even-cat fidelity landscapes for homodyne-bred resources with $N=$(a)2, (b)4, (c)8, (d)16 squeezed single-photon seeds. Each panel shows $F^{(N)}_+(\sigma_{\mathrm{eff}},\alpha)$ versus seed squeezing (dB) and target amplitude $\alpha$. The dashed curve denotes the maximizing ridge $\sigma_{\mathrm{eff}}^*(\alpha)$, the solid red contour marks $F^{(N)}_+=0.99$, and the cross indicates the largest-amplitude point satisfying this threshold, $\alpha^{(N)}_{\mathrm{max}}$.}
\label{fig2_landscape}
\end{figure}

Fig.~\ref{fig2_landscape} visualizes the same optimization in terms of the seed squeezing, expressed in dB, and the target amplitude $\alpha$, with $\sigma_{\mathrm{eff}}=e^{2r}$. As $N$ increases from $2$ to $16$, this threshold amplitude grows from $\alpha^{(2)}_{\mathrm{max}}\simeq1.65$ to $2.38$, $3.31$, and $\alpha^{(16)}_{\mathrm{max}}\simeq4.65$, which is clearly shown in Fig.~\ref{fig1_overview}(c). At the same time, the high-fidelity region expands over a broader range of target amplitudes, indicating that larger cat states are attainable. The corresponding squeezing lies in the range $1.7 \lesssim r_\mathrm{dB} \lesssim 4.3$, where $r_\mathrm{dB}=20r/\ln 10$, indicating that the required squeezing is relatively modest.

The growth of $\alpha^{(N)}_{\mathrm{max}}$ is close to a factor of $\sqrt{2}$ per breeding round, consistent with the amplitude scaling of balanced cat breeding~\cite{vasconcelos_all-optical_2010, lund_conditional_2004}. A balanced beam splitter rescales coherent amplitudes by $1/\sqrt{2}$; adding one round therefore increases the attainable output amplitude by roughly $\sqrt{2}$. This scaling identifies the breeding rounds as the amplitude-amplification stage of the protocol, while the squeezed single-photon states supply the initial non-Gaussian resource.

In the symmetric circuit, the mapping $\sigma_{\mathrm{eff}}=e^{2r}$ gives a direct inverse-design relation between the effective width and the seed squeezing. Thus any chosen effective width corresponds to
\begin{equation}
    r=\frac{1}{2}\ln\sigma_{\mathrm{eff}}.
\end{equation}
The maximizing ridge in Fig.~\ref{fig2_landscape} specifies the fidelity-optimized point on this relation for each target amplitude. This relation can be extended to a target squeezed cat state $\ket{\mathrm{SCS}_+(\alpha,\zeta)}$. The squeezing operation $\hat{S}(\zeta)$ preserves the polynomial order and rescales the width from $\sigma_{\mathrm{eff}}$ to $\sigma_{\mathrm{eff}}e^{-2\zeta}$. This rescaling is equivalent to a uniform shift in the common seed squeezing from $r$ to $r-\zeta$. Hence, after choosing the plain-cat optimum $\sigma_{\mathrm{eff}}$, the corresponding squeezed-cat target is obtained by applying the same shift to the seed squeezing, without changing the circuit topology.

The simple shift $r\to r-\zeta$ is a consequence of the symmetric breeding construction, rather than a generic property of Gaussian circuits. For a general multimode Gaussian circuit, propagating an output squeezing backward to the seed level would generally result in multiple input squeezings and passive rotations in the Bloch–Messiah decomposition. In the present construction, however, the identical seeds and the preservation of $\sigma_{\mathrm{eff}}$ at every breeding round reduce this redistribution to a single common change of the seed squeezing. Once $N$, $\alpha$, and the desired output squeezing $\zeta$ are fixed, the required common seed squeezing is therefore determined directly.

\section{GKP breeding \label{GKP_generation}}

The analysis above treated the homodyne-bred polynomial--Gaussian state as a cat-state approximation, quantifying its fidelity to a target cat state and the seed squeezing required to reach a given operating point. In breeding-based optical generation of GKP states, cat states serve as intermediate resources~\cite{vasconcelos_all-optical_2010,weigand_generating_2018}. Existing schemes typically assume ideal squeezed cat states with prescribed amplitude and squeezing as inputs, whereas the resource available here is the polynomial--Gaussian state $\psi^{(N)}_{\mathrm{out}}$ of Eq.~\eqref{N_output}, generated from squeezed single-photon seeds. We therefore use the inverse-design relation derived above to fix the input resource, and then ask how the finite cat-resource size determines the stabilizer squeezing, lattice matching, and resource cost of the GKP state produced in the subsequent breeding stage.

\subsection{GKP state definition and conventions}

Before turning to the numerical GKP-breeding results, we first fix the lattice convention. A single-mode rectangular GKP qubit code is defined by the joint $+1$ eigenspace of the displacement stabilizers $\hat{\mathcal{S}}_x=e^{-iD_x\hat{p}}$ and $\hat{\mathcal{S}}_p=e^{iD_p\hat{x}}$, with $D_xD_p=4\pi$~\cite{gottesman_encoding_2001}. Here, $D_x$ and $D_p$ specify the code-stabilizer displacement periods. We separately use $d_x$ and $d_p$ for the primitive periods of the individual grid states considered below. For an unshifted comb, these satisfy $d_xd_p=2\pi$. The corresponding ideal grid state in the $\hat{x}$-quadrature representation takes the form
\begin{equation}
    \psi^{\mathrm{GKP}}_{\mathrm{ideal}}(x)\propto \sum_{j\in\mathbb{Z}}\delta(x-jd_x) ,
\end{equation}
where $\delta\left(\cdot\right)$ denotes the Dirac delta function. This ideal state cannot be prepared physically: its peaks are infinitely squeezed, and the comb extends over infinitely many lattice sites without a decaying envelope, leading to divergent energy.

Thus, we compare the generated states with finite-energy approximate GKP states~\cite{gottesman_encoding_2001, vasconcelos_all-optical_2010}. In this approximation, each Dirac peak is replaced by a Gaussian peak of width $\Delta$, and the comb is multiplied by a Gaussian envelope controlled by $\kappa$. Following this standard Gaussian-comb form, we use the $\hat{x}$-quadrature fitting wavefunction:
\begin{equation}
    \psi^{\mathrm{GKP}}_{\mathrm{approx}}(x;\Delta, \kappa)=\frac{1}{\sqrt{\mathcal{N}_{\Delta, \kappa}}}\sum_{j\in\mathbb{Z}}c_jg_{\Delta}(x-jd_x)
\end{equation}
where $c_j=e^{-\kappa^2(jd_x)^2/2},\quad g_{\Delta}(x)=e^{-x^2/2\Delta^2}$, $d_x$ is set by the target GKP lattice, and $\sqrt{\mathcal{N}_{\Delta, \kappa}}$ is the normalization constant.
In the following analysis, we use the square-lattice qunaught state $\ket{\varnothing}$, whose primitive lattice periods are $d_x=d_p=\sqrt{2\pi}$. Although the qunaught state does not encode a logical qubit, two ideal qunaught states mixed on a balanced beam splitter produce a GKP Bell state~\cite{aghaee_rad_scaling_2025, walshe_continuous_2020}. The ideal qunaught state is related to the logical GKP state $\ket{0_L}$ by a single-mode squeezing operation, $\ket{0_L}\propto\hat{S}\left(-\ln{\sqrt{2}}\right)\ket{\varnothing}$, which transforms the lattice periods from $\left(\sqrt{2\pi},\sqrt{2\pi}\right)$ to $\left(2\sqrt{\pi},\sqrt{\pi}\right)$. We therefore adopt the qunaught state as the common reference for the performance analysis below.

With this convention, the ideal GKP limit is approached as $\Delta, \kappa \rightarrow 0$. With the finite-energy reference fixed, we now introduce the figures of merit used to assess the generated states. The optimized overlap with the reference family is defined as
\begin{equation}
\label{finite_GKP_fidelity}
    F_\mathrm{GKP} = \max_{\Delta,\kappa} \braket{\psi^{\mathrm{GKP}}_{\mathrm{approx}}|\rho|\psi^{\mathrm{GKP}}_{\mathrm{approx}}}.
\end{equation}
A large value of $F_\mathrm{GKP}$ indicates that the generated state $\rho$ is well described by the chosen finite-energy grid state. The fitted $\Delta$ and $\kappa$, however, are parameters of this reference family, and $F_\mathrm{GKP}$ alone does not determine the operational GKP quality. Hence, we use stabilizer-based effective squeezing as the primary figure of merit, while retaining $F_\mathrm{GKP}$ as a structural diagnostic.

For a state $\rho$ and a target lattice with periods $d_x$ and $d_p$, we denote the corresponding state-stabilizing displacements by $\hat{\mathcal{T}}_x=e^{-id_x\hat{p}}$ and $\hat{\mathcal{T}}_p=e^{id_p\hat{x}}$. The effective squeezing in the two quadratures is then extracted from their expectation values,
\begin{equation}
\begin{aligned}
    & \Delta^{\mathrm{eff}}_x = \frac{1}{d_p}\sqrt{-2\ln|\mathrm{Tr}(\rho\,\hat{\mathcal{T}}_p)|^2} , \\
    & \Delta^{\mathrm{eff}}_p = \frac{1}{d_x}\sqrt{-2\ln|\mathrm{Tr}(\rho\,\hat{\mathcal{T}}_x)|^2} .
\end{aligned}
\end{equation}
We can express the effective squeezing in decibel units as $\Delta^{\mathrm{eff}}_{\mathrm{dB}}=-20\log_{10}\Delta^{\mathrm{eff}}$. We also use the symmetric effective squeezing, defined by combining the two quadrature variances as~\cite{aghaee_rad_scaling_2025}
\begin{equation}
    (\Delta^{\mathrm{eff}}_{\mathrm{sym}})^2=\frac{(\Delta^{\mathrm{eff}}_x)^2+(\Delta^{\mathrm{eff}}_p)^2}{2}.
\end{equation}
This stabilizer-based characterization ties the quality of the state to displacement moments rather than to the overlap with a particular reference state, in line with recent nonlinear-squeezing approaches to GKP and related grid-state resources~\cite{marek_ground_2024, kuchar_nonlinear_2025}.

\subsection{Output-state characterization}

\begin{table}[t]
\caption{One-step GKP generation from the bred resource, evaluated with respect to the square-lattice qunaught state. The seed squeezing denotes the common squeezing applied to the single-photon seeds used to prepare the GKP input resources, with the lattice-matching shift incorporated through the inverse-design relation. The resource cost $2N$ is the number of squeezed single-photon seeds.\\}
\label{one_step_table}
\renewcommand{\arraystretch}{1.3}
\setlength{\tabcolsep}{5.5pt}
\begin{ruledtabular}
\begin{tabular}{ccccc}
\multirow{2}{*}{\raisebox{-4ex}{\shortstack{Resource cost \\[3pt] $2N$}}} & \multirow{2}{*}{\raisebox{-5ex}{\shortstack{Seed squeezing \\[3pt] (dB)}}} & \multicolumn{3}{c}{Effective squeezing (dB)} \\[3pt]
\cmidrule(lr){3-5}
\noalign{\vskip 3pt}
 &  & $\Delta^{\mathrm{eff}}_{x}$ & $\Delta^{\mathrm{eff}}_{p}$ & $\Delta^{\mathrm{eff}}_{\mathrm{sym}}$ \\[3pt]
\hline
\noalign{\vskip 3pt}
$4$  & $1.35$ & $1.504$  & $6.432$ & $3.304$ \\
$8$  & $-1.40$ & $4.521$  & $5.960$ & $5.181$ \\
$16$ & $-4.24$ & $7.368$  & $5.879$ & $6.560$ \\
$32$ & $-7.16$ & $10.232$ & $5.880$ & $7.532$
\end{tabular}
\end{ruledtabular}
\end{table}

\begin{figure*}
\centering
\includegraphics[width=0.99\textwidth]{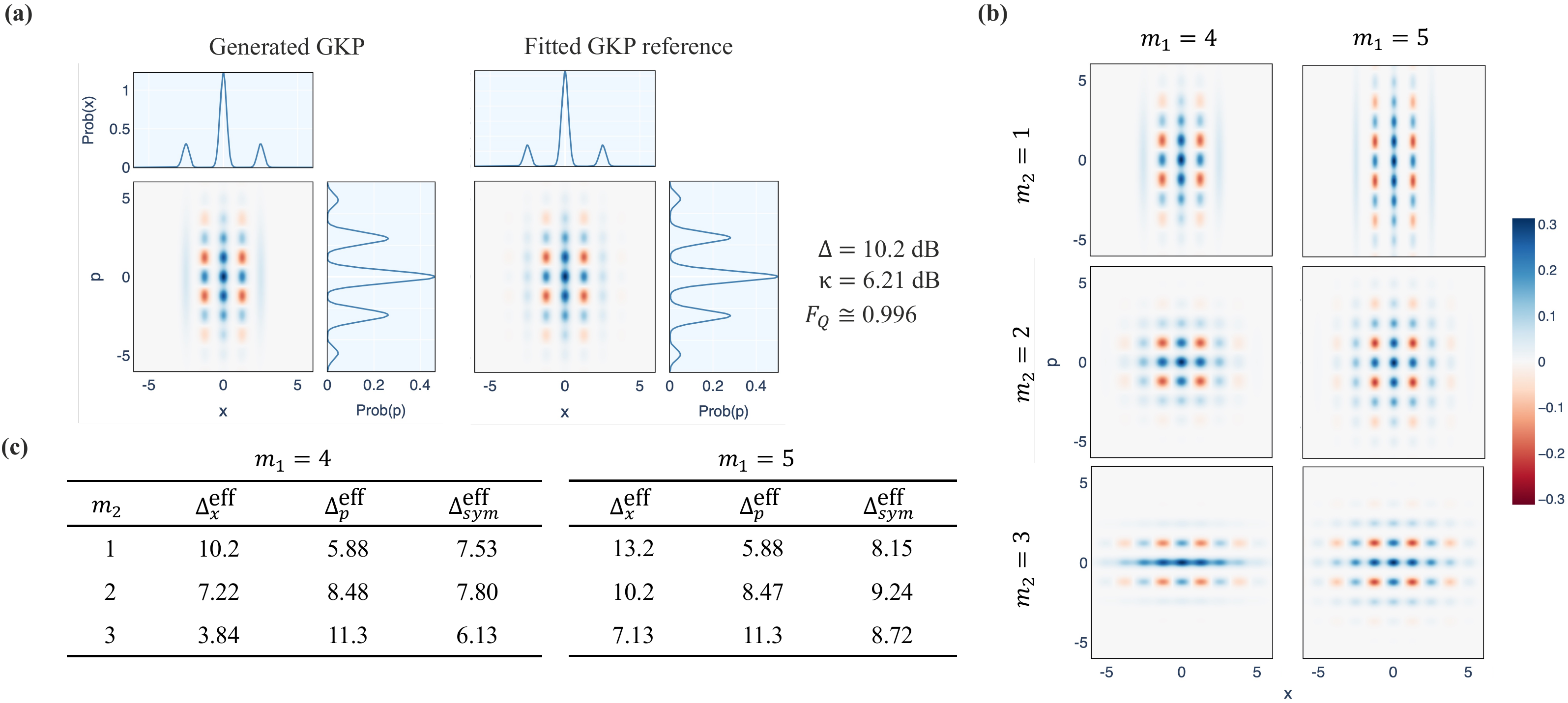}
\caption{GKP-to-GKP breeding from the homodyne-bred cat resources. (a) Representative one-step GKP output and the corresponding optimized finite-energy GKP state from the $F_\mathrm{GKP}$ fit, together with their $\hat{x}$- and $\hat{p}$-marginal distributions.  (b) Wigner functions of $m_2$-GKP states generated from $m_1$-cat resources, with the final lattice fixed to the qunaught state $d_x=d_p=\sqrt{2\pi}$. (c) Effective squeezings for the states in (b). All values are given in dB.
}
\label{fig3_gkp}
\end{figure*}

As a baseline case, we first consider one-step GKP generation. Two identical copies of the cat-bred resource state $\psi^{(N)}_{\mathrm{out}}$, with $N=2^{m_1}$, are mixed on a balanced beam splitter, and one output mode is conditioned on the homodyne outcome $p=0$. With the qunaught lattice fixed as above, the resource squeezing $\zeta$ is chosen so that the peak spacing after the balanced fusion matches the target lattice period, $d_x=2\alpha e^{-\zeta}$, giving $\zeta=\ln(2\alpha/d_x)$. Through the inverse-design relation derived above, this resource squeezing can equivalently be incorporated into the squeezing applied to the initial single-photon seeds. The amplitudes are taken at the high-fidelity operating points of the cat-breeding stage.

All one-step outputs considered here have $F_\mathrm{GKP}>0.99$, indicating close agreement with their optimized finite-energy reference states. Fig.~\ref{fig3_gkp}(a) illustrates this agreement for a representative one-step output, together with the corresponding $\hat{x}$- and $\hat{p}$-marginal distributions. Table~\ref{one_step_table} lists the initial seed squeezing in dB and the resulting stabilizer-based effective squeezings. At fixed qunaught lattice spacing, increasing $N$ primarily improves $\Delta^{\mathrm{eff}}_x$, expressed in dB, while the corresponding value $\Delta^{\mathrm{eff}}_p$ varies much less. A larger cat resource thus improves the effective squeezing primarily along the $\hat{x}$ quadrature, rather than comparably in both quadratures. As a result, the one-step GKP state remains anisotropic. 

The remaining question is how the anisotropic effective squeezing of one-step outputs can be used when the final target lattice is kept fixed. To examine this within the existing GKP-breeding framework, we consider additional $m_2-1$ layers of GKP-to-GKP breeding, with all final outputs matched to the qunaught lattice. Each layer applies the same balanced beam splitter and homodyne conditioning operation illustrated in Fig.~\ref{fig1_overview}(b) to two copies of the output from the preceding layer. As shown in Fig.~\ref{fig3_gkp}(b), increasing $m_2$ increases the number of resolved grid peaks in the Wigner function, thereby extending the finite comb further in phase space. This extension is the main structural advantage of the GKP-to-GKP step. The effective squeezings clarify the role of the added GKP-to-GKP layers. For a fixed $m_1$-cat resource, the increase in $\Delta_p^{\mathrm{eff}}$ is accompanied by a reduction in $\Delta_x^{\mathrm{eff}}$, rather than an improvement in both quadratures simultaneously. Hence, increasing $m_2$ trades effective squeezing between the two quadratures, allowing the anisotropy of the one-step output to be reduced. Fig.~\ref{fig3_gkp}(c) shows the effective squeezing obtained, indicating that $m_2=2$ yields the most symmetric GKP output state for both $m_1=4$ and $m_1=5$. Each additional GKP-to-GKP layer consumes two copies of the previous GKP state and therefore doubles the number of squeezed single-photon seeds required to prepare one output state. In this sense, GKP-to-GKP breeding improves the finite-grid structure and redistributes the effective squeezing between the two quadratures, at the cost of a larger number of squeezed single-photon seeds.

\subsection{Approximate stellar rank}

To make a resource-theoretic analysis of our scheme, we consider the approximate stellar rank $r^*_\epsilon$~\cite{hahn_assessing_2026}, which quantifies the non-Gaussianity of a quantum state in terms of the minimum Fock-state excitation number required to approximate it. The approximate stellar rank is defined as the minimum stellar rank~\cite{chabaud_stellar_2020} among states having fidelity at least $1-\epsilon$ with a target state. For each one-step GKP output, we take the finite-energy qunaught reference that maximizes $F_\mathrm{GKP}$ in Eq.~\eqref{finite_GKP_fidelity} and set $1-\epsilon=F_\mathrm{GKP}$, so that $r^*_{1-F_\mathrm{GKP}}$ gives the minimum stellar rank required to approximate the same reference at the fidelity already attained by the protocol.

\begin{figure}[t]
\includegraphics[width=\linewidth]{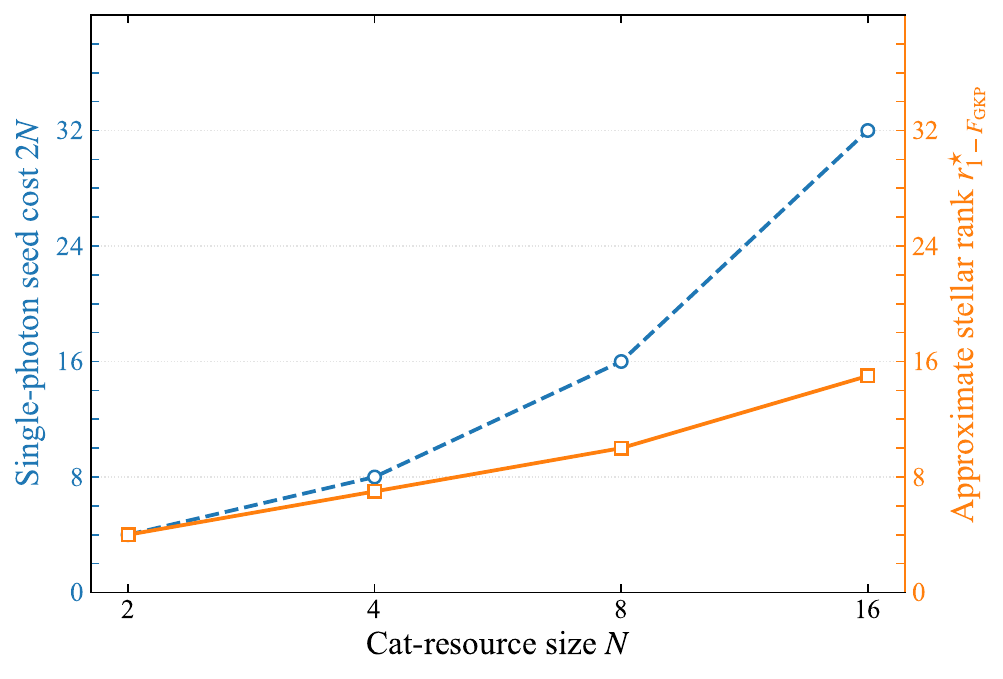}
\caption{Comparison of the physical single-photon resource cost $2N$ and the corresponding approximate stellar rank $r^*_{1-F_\mathrm{GKP}}$ for the one-step GKP outputs generated from $N=2,4,8,$ and $16$ homodyne-bred cat resources. The approximate stellar rank is evaluated at the achieved fidelity $1-\epsilon=F_\mathrm{GKP}$.}
\label{fig4_approxstellarrank}
\end{figure}

Since each squeezed single-photon seed has stellar rank one, the additivity and monotonicity of stellar rank imply the lower bound $r^*_{1-F_\mathrm{GKP}}$ on the required number of such rank-one non-Gaussian inputs. For physical resource costs $2N=4,8,16,$ and $32$, the optimized overlaps are $F_\mathrm{GKP}=0.9966,0.9957,0.9956,$ and $0.9958$, respectively. The corresponding approximate stellar ranks are $r^*_{1-F_\mathrm{GKP}}=4,7,10,$ and $15$, as shown in Fig.~\ref{fig4_approxstellarrank}. Thus, the stellar-rank bound is saturated for $2N=4$ and remains close to the physical cost for $2N=8$, whereas the separation becomes more pronounced for $2N=16$ and $32$. The smaller breeding trees therefore operate close to the stellar-rank lower bound on the required resource, while the physical seed cost increasingly exceeds this bound as the breeding tree is scaled up.

\section{Finite-window analysis}
\label{finite_window}
The analysis of Sec.~\ref{breeding_and_resource} was carried out at the ideal homodyne outcome $p=0$, where the heralded output state takes the closed monomial--Gaussian form and the even-cat fidelity and optimal seed squeezing follow analytically. In practice, a continuous-variable homodyne measurement returns an exact quadrature value with vanishing probability, so any realistic implementation must accept outcomes within a finite acceptance window $|p_i|\leq w$ around the target~\cite{laghaout_amplification_2013}. The width of this window sets a trade-off intrinsic to homodyne heralding. Widening it raises the success probability, but at the same time admits outcomes that deviate further from the ideal $p=0$ output and can lower the fidelity of the heralded state.

To characterize this trade-off, we need the heralded state and its norm as functions of the full outcome vector $\mathbf{p}=(p_1,\dots,p_{N-1})$. For squeezed-cat inputs, related multivariate-Gaussian methods have been used to propagate the beam-splitter cascade and evaluate the dependence of the heralded state on nonzero homodyne outcomes~\cite{solodovnikova_loss_2025}. Here we apply the same multivariate-Gaussian reduction to the squeezed-single-photon breeding path of Sec.~\ref{breeding_and_resource}. Tracking the conditional polynomial round by round grows rapidly more involved as $N$ increases, so we instead incorporate the final $\pi/2$ phase-space rotation into the measurement step from the outset, treating the entire circuit as a single $N$-dimensional Gaussian Fourier transform.

The corresponding calculation is given in Appendix~\ref{AppendixB}. For a homodyne outcome vector $\mathbf{p}=(p_1,\dots,p_{N-1})$, the unnormalized heralded state is
\begin{equation}
\begin{aligned}
    \tilde{\psi}_N(x;\mathbf{p})&\propto Q_N(x;\mathbf{p})\exp(-\frac{x^2+|\mathbf{p}|^2}{2\sigma_{\mathrm{eff}}}),\\
    \quad Q_N(x;\mathbf{p})&=N^{N/2}\prod^N_{j=1}(O^T_N\mathbf{q})_j,
\end{aligned}
\end{equation}
where $O_N$ is the real orthogonal matrix of the balanced beam-splitter tree and $\mathbf{q}=(x,p_1,\dots,p_{N-1})^T$ collects the output quadrature and the homodyne outcomes. $Q_N$ is the outcome-dependent polynomial that replaces the ideal monomial $x^N$ and the $x$-dependent Gaussian $e^{-x^2/2\sigma_{\mathrm{eff}}}$ is the same as in Eq.~\eqref{N_output}. The additional factor $e^{-|\mathbf{p}|^2/2\sigma_{\mathrm{eff}}}$ depends only on the outcomes and contributes to the outcome probability density. At $\mathbf{p}=\mathbf{0}$ the polynomial recovers the ideal output $Q_N(x;\mathbf{0})=x^N$, while nonzero outcomes deform it away from the ideal monomial and, for higher-order breeding circuits, can also break its exact even parity. These polynomial deformations become stronger as the accepted outcomes move farther from zero and are the origin of the fidelity loss once a finite window is opened. The probability density of the outcome is obtained from the norm of the unnormalized state,
\begin{equation}
\begin{aligned}
    P_N(\mathbf{p})&\propto e^{-|\mathbf{p}|^2/\sigma_{\mathrm{eff}}}\mathcal{N}_N(\mathbf{p}),\\[4pt]
    \mathcal{N}_N(\mathbf{p})&=\int^\infty_{-\infty}dx|Q_N(x;\mathbf{p})|^2e^{-x^2/\sigma_{\mathrm{eff}}} .
\end{aligned}
\end{equation}
The conditional fidelity between the target even cat and the normalized state $\psi_N(\mathbf{p})$ is
\begin{equation}
    F_N(\mathbf{p})=|\braket{\mathrm{Cat}_+(\alpha)|\psi_N(\mathbf{p})}|^2 .
\end{equation}
Because $Q_N$ is a finite polynomial, both $F_N(\mathbf{p})$ and $P_N(\mathbf{p})$ are finite sums of Gaussian moments and are obtained in closed form. 

A finite window accepts every outcome inside the box $\mathcal{W}(w)=[-w,w]^{N-1}$, and we characterize it by two quantities,
\begin{equation}
\begin{aligned}
    P^{(N)}_{\mathrm{acc}}(w)&=\frac{\int_{\mathcal{W}(w)}d^{N-1}\mathbf{p}\,e^{-|\mathbf{p}|^2/\sigma_{\mathrm{eff}}}\mathcal{N}_N(\mathbf{p})}{\int_{\mathbb{R}^{N-1}}d^{N-1}\mathbf{p}\,e^{-|\mathbf{p}|^2/\sigma_{\mathrm{eff}}}\mathcal{N}_N(\mathbf{p})},\\[4pt]
    \bar{F}_N(w)&=\frac{\int_{\mathcal{W}(w)}d^{N-1}\mathbf{p}\,P_N(\mathbf{p})F_N(\mathbf{p})}{\int_{\mathcal{W}(w)}d^{N-1}\mathbf{p}\,P_N(\mathbf{p})}.
\end{aligned}
\end{equation}
The success probability $P^{(N)}_{\mathrm{acc}}(w)$ is the fraction of the outcome distribution falling inside the window, and the window-averaged fidelity $\bar{F}_N(w)$ is the conditional fidelity averaged over the accepted region. In both cases, the closed-form expressions are integrated over the continuous box, rather than estimated from a sampled set of outcomes.

\begin{figure}
\includegraphics[width=\linewidth]{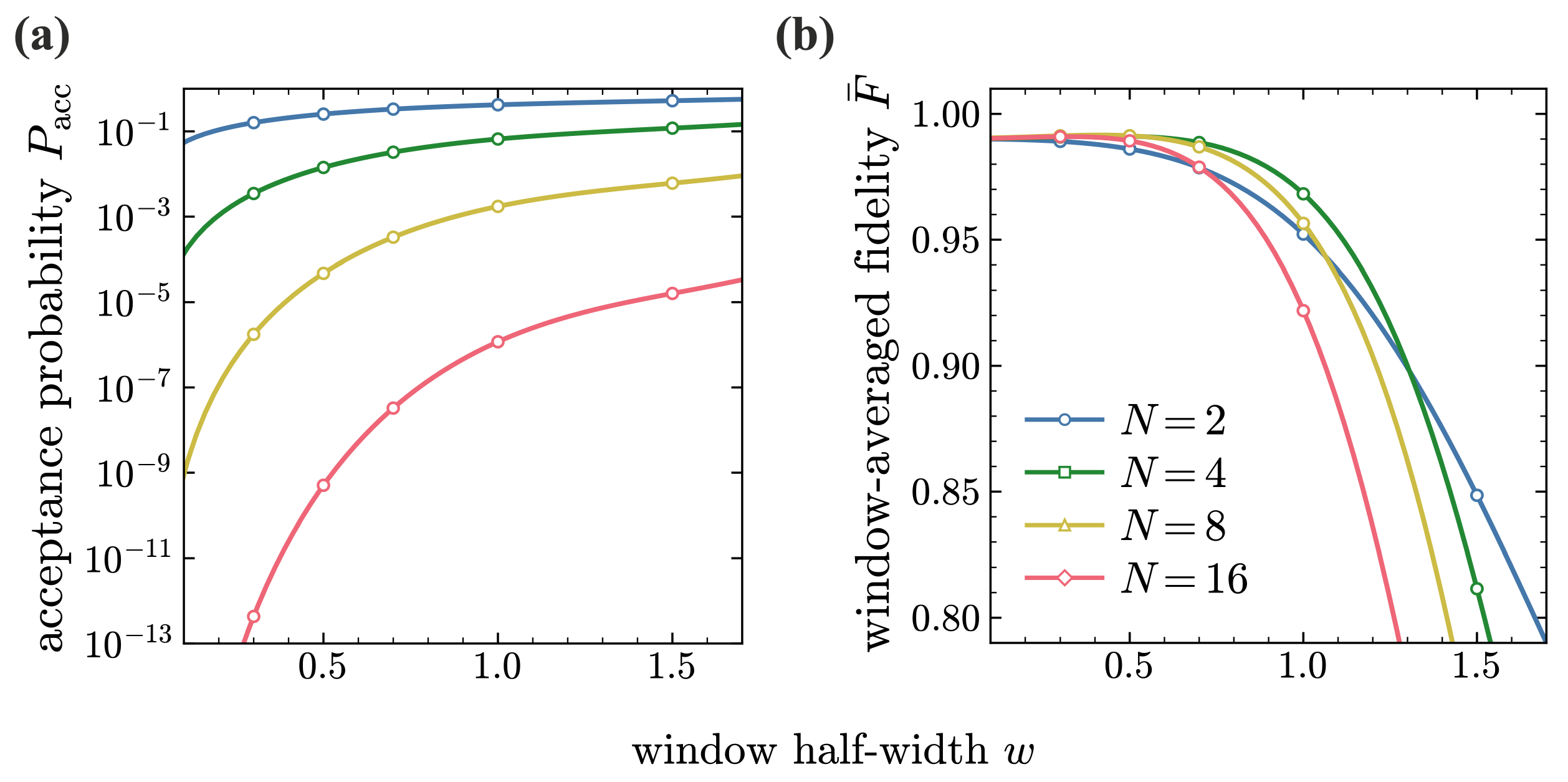}
\caption{Plot illustrating (a) success probability $P^{(N)}_{\mathrm{acc}}(w)$ and (b) window-averaged fidelity $\bar{F}_N(w)$ for homodyne-bred resources with $N=2,4,8,16$ squeezed single-photon seeds, as a function of the acceptance-window half-width $w$. We evaluate the success probability and the fidelity at the operating point fixed in Sec.~\ref{breeding_and_resource} for each $N$: the target amplitude $\alpha^{(N)}_{\mathrm{max}}$ and the effective width $\sigma^*_{\mathrm{eff}}$ that maximizes the ideal even-cat fidelity. All values are obtained analytically from the finite-outcome state, without sampling.}
\label{fig5_finitewindow}
\end{figure}

We evaluate the success probability and the fidelity at the operating point fixed in Sec.~\ref{breeding_and_resource} for each $N$. The resulting trade-off is summarized in Fig.~\ref{fig5_finitewindow}, which shows $P^{(N)}_{\mathrm{acc}}(w)$ and $\bar{F}_N(w)$ for $N=2,4,8,16$ over a range of window half-widths. One might expect the finite-window fidelity to degrade progressively with $N$, since a larger resource requires more conditioning measurements. The averaged fidelity does not follow this expectation. Up to intermediate windows $w\lesssim1.0$ all four resources remain above $0.9$, and over part of this range the larger resources retain a fidelity comparable to, or slightly higher than, the $N=2$ resource. The ordering becomes more systematic only once the fidelity begins to fall steeply: beyond $w\simeq1.0$ the larger resources degrade more rapidly. Thus, the finite-window degradation is not set by $N$ alone but by how far the accepted outcomes deform the finite-outcome polynomial $Q_N(x;\mathbf{p})$ away from the ideal monomial.

This locates the useful operating regime at an intermediate window rather than the infinitesimal-window limit. Near $w\simeq1.0$ the averaged fidelity stays above $0.9$ for all $N$ while the acceptance probability remains finite, ranging from $4.20\times10^{-1}$ at $N=2$ down to $1.18\times10^{-6}$ at $N=16$. The generation rate associated with this probability also depends on the delivery rate of complete seed sets, as discussed in Sec.~\ref{comparison}. The finite-window analysis therefore refines the ideal $\mathbf{p}=\mathbf{0}$ result: increasing $N$ extends the attainable cat amplitude, which in turn improves the one-step GKP resource of Sec.~\ref{GKP_generation}, but the usable window is bounded from both sides, with $w\simeq0.7$--$1.0$ providing the practical compromise.

\section{Loss tolerance}
\label{loss_tolerance}
We now examine the response of the protocol to photon loss. Because the breeding protocol is symmetric, losses arising from the beam splitters and detectors can be commuted to the input and described by a single effective loss parameter~\cite{solodovnikova_loss_2025}. In our case, this means that each squeezed single-photon seed is passed through the same pure-loss channel $\mathcal{L}_\eta$, with the transmissivity $\eta$, immediately after preparation and before any beam-splitter interaction. A single cat resource is generated from
\begin{equation}
    \left[\mathcal{L}_\eta(\hat{S}(r)\ket{1}\bra{1}\hat{S}^\dagger(r))\right]^{\otimes N} .
\end{equation}
The resulting lossy seeds are propagated through the beam-splitter cascade and the $\mathbf{p}=\mathbf{0}$ homodyne-heralding, so that all quantities reported for the cat resources are evaluated on the heralded output state. Two copies of the resulting lossy squeezed-cat resource are then used as the inputs to the one-step GKP breeding stage. Thus the GKP metrics characterize states bred from loss-degraded seeds propagated through the entire protocol, rather than from an ideal resource followed by output loss.

Since the loss channel does not commute with squeezing, we compare two approaches. In the first, we generate the cat state and then apply squeezing to match the GKP lattice spacing. Then, the plain-cat fidelity $F_{\mathrm{cat}}$ measures the overlap of the heralded lossy output with $\ket{\mathrm{Cat}_+(\alpha^{(N)}_{\mathrm{max}})}$. In the second, we apply all squeezing to the seed state, yielding a spacing-matched squeezed cat state $\ket{\mathrm{SCS}_+(\alpha^{(N)}_{\mathrm{max}},\zeta_N)}$. The squeezed-cat fidelity $F_{\mathrm{SCS}}$ therefore measures the overlap with this GKP input resource, and the same lossy squeezed cat state is used for the GKP breeding.

\begin{figure}
\includegraphics[width=\linewidth]{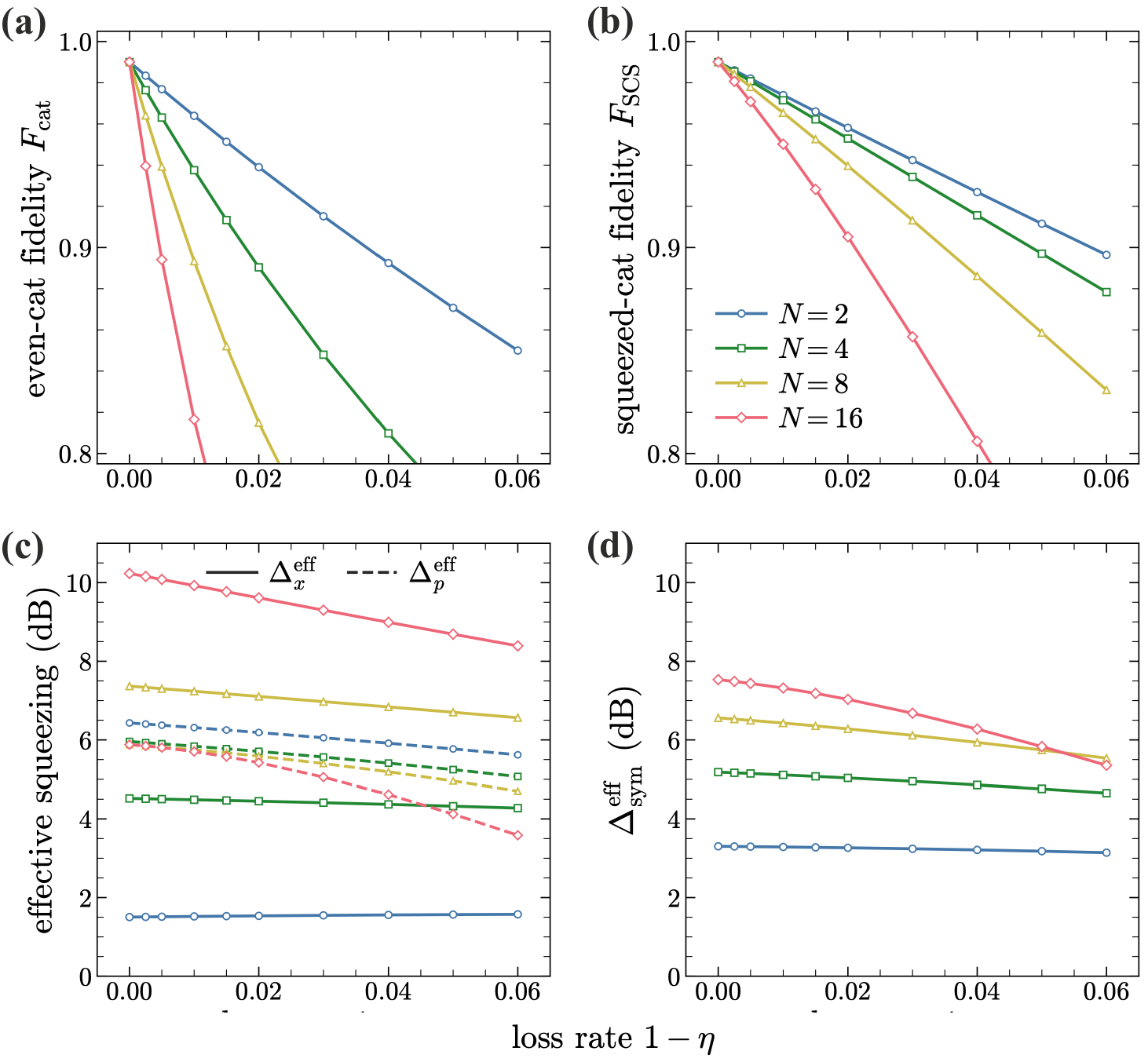}
\caption{Photon loss response of the bred cat and one-step GKP resources for $N=2,4,8,16$ squeezed single-photon seeds, as a function of the loss rate $1-\eta$. The figure shows (a) the plain even-cat fidelity $F_{\mathrm{cat}}$, (b) the squeezed-cat fidelity $F_{\mathrm{SCS}}$ of the spacing-matched resource used for GKP breeding, (c) the effective squeezings $\Delta_x^{\mathrm{eff}}$(solid) and $\Delta_p^{\mathrm{eff}}$(dashed) of the one-step GKP output, and (d) the corresponding symmetric effective squeezing $\Delta_{\mathrm{sym}}^{\mathrm{eff}}$. The effective squeezings are expressed in dB.}
\label{fig6_loss}
\end{figure}

Fig.~\ref{fig6_loss}(a,b) show that the bare cat diagnostic and the GKP-relevant resource respond differently to seed loss. The plain-cat fidelity decreases rapidly with increasing $N$, as expected for larger-amplitude cats whose coherence is more vulnerable to photon loss. The squeezed-cat fidelity, however, degrades more gradually. This indicates that the spacing-matching approach, which is actually required for GKP generation, is more stable than the plain-cat approach.

We therefore employ the cat resource states bred with the space-matching approach as inputs to GKP breeding and analyze the stabilizer metrics of the one-step GKP output, shown in Fig.~\ref{fig6_loss}(c,d). Loss reduces the effective squeezing, but the output retains finite squeezing rather than degrading rapidly under small amounts of seed loss. The main limitation appears in the asymmetric behavior of the degradation. Larger resources give higher $\Delta_x^{\mathrm{eff}}$ in the lossless limit, but they are also more sensitive to loss. This effect is most pronounced for resources with large $N$. While the $N=16$ resource starts with the largest $\Delta_x^{\mathrm{eff}}$, its $\Delta_p^{\mathrm{eff}}$ decreases much faster than that of the smaller resources. As a result, $N=8$ achieves larger effective squeezing than $N=16$ when $ \eta \lesssim 0.95 $. This behavior is consistent with the amplitude dependence of cat decoherence: for a cat of amplitude $\alpha$, pure loss suppresses the off-diagonal coherence on the scale $\exp\left[-2(1-\eta)\alpha^2\right]$~\cite{zhang_quantifying_2021}. Since the $N=16$ operating point has the largest $\alpha$, the same loss percentage corresponds to a larger decoherence parameter $(1-\eta)\alpha^2$. The effect is especially visible in
$\Delta_p^{\mathrm{eff}}$, which is obtained from $\mathrm{Tr}[\rho e^{-id_x\hat{p}}]$ and is therefore sensitive to coherence between neighboring grid peaks.

\section{Comparison with previous works}
\label{comparison}
The breeding path analyzed in this work occupies a specific position among optical non-Gaussian state generators, and it is useful to make that position explicit. Homodyne breeding itself is an established construction~\cite{vasconcelos_all-optical_2010, weigand_generating_2018}. In the photon-subtraction family, including generalized photon subtraction (GPS)~\cite{takase_generation_2021}, Gaussian inputs are combined on a beam splitter and a PNR measurement heralds the non-Gaussian output. The same principle underlies the GKP synthesizer of Takase~\textit{et al.}~\cite{takase_gottesman-kitaev-preskill_2023}. In both cases the preparation is heralded and probabilistic, with the selected photon-number outcome determining the conditional non-Gaussian state.

For photon-counting implementations, the relevant detector characteristics include photon-number resolution, detection efficiency, and repetition rate. Current detector technologies offer different trade-offs among these characteristics. Transition-edge sensors (TESs) offer high detection efficiency and reliable photon-number resolution, but their thermal recovery limits the repetition rate. Machine learning readout has enabled operation at $800\,\mathrm{kHz}$ with accurate assignment up to at least five photons~\cite{li_boosting_2025}. Superconducting-nanowire detectors occupy the complementary regime. They offer fast, efficient single-photon detection and can provide photon-number resolution through spatial multiplexing. For example, a scalable parallel SNSPD reported two- and three-photon efficiencies of about $75\%$ and $62\%$, respectively. In this architecture, finite detection efficiency and multiple photons reaching the same pixel reduce the probability of correctly resolving the incident photon numbers~\cite{stasi_enhanced_2025}. Although a trial repetition rate of $100\,\mathrm{MHz}$ has been assumed in GPS generation-rate estimates~\cite{takase_generation_2021}, a high output rate also requires efficient photon-number discrimination at the required photon number.

With clocked squeezed-single-photon seeds as the input resource, the present implementation avoids high-$n$ PNR heralding in cat preparation and GKP breeding. The trial repetition rate therefore depends on the delivery of complete seed sets and the homodyne readout. A quantum-dot source has demonstrated a zero-delay second-order correlation $g^{(2)}(0)=0.0205$, indistinguishability $0.986$, and system efficiency $0.712$~\cite{ding_high-efficiency_2025}. Quantum-dot platforms have also operated at an excitation clock of $80\,\mathrm{MHz}$~\cite{maring_versatile_2024, hauser_deterministic_2026}. Integrated balanced homodyne receivers have demonstrated a shot-noise-limited bandwidth exceeding $20\,\mathrm{GHz}$~\cite{bruynsteen_integrated_2021}, supporting fast quadrature readout without photon-number discrimination. The practical requirements are shifted to synchronized delivery of many identical seeds, mode matching, and control of photon loss.

For the probability comparison, GPS conditions on selected photon-counting outcomes, whereas the present protocol uses a finite homodyne acceptance window. At the cat level, we compare the optimized GPS probabilities of Ref.~\cite{takase_generation_2021} with our finite-window results at $w=1.0$, using $n\sim\alpha^2$ as a rough photon-number scale. In this representative comparison, homodyne breeding is more probable at small amplitude $\alpha=1.65$ and $2.38$ ($\alpha\lesssim3$). At $\alpha=3.31$, the breeding probability lies between the GPS results for $7$ and $10\,\mathrm{dB}$ input squeezing. At $\alpha=4.65$, it falls below the GPS results for $7$--$10\,\mathrm{dB}$ input squeezing. The crossover therefore depends on the GPS input squeezing. These probabilities are evaluated under the respective protocol assumptions: optimized Gaussian inputs for GPS and supplied squeezed-single-photon seeds for homodyne breeding.

At the GKP level, our ideal $2N=16$ one-step state reaches a symmetric effective squeezing of $6.56\,\mathrm{dB}$. This state is formed by mixing two independently bred $N=8$ cat resources on a balanced beam splitter and conditioning the resulting mode. Our finite-window analysis in Sec.~\ref{finite_window} gives $[P^{(8)}_{\mathrm{acc}}(1.0)]^2\simeq3.0\times10^{-6}$ for preparing the two input cats. This probability excludes the subsequent GKP-fusion conditioning; the $6.56\,\mathrm{dB}$ output squeezing is evaluated separately at the ideal conditioning point of Sec.~\ref{GKP_generation}. For reference, the synthesizer of Ref.~\cite{takase_gottesman-kitaev-preskill_2023} reports a $6.4\,\mathrm{dB}$ target with a success probability of $1.06\times10^{-3}$. These examples provide benchmarks at similar reported squeezing levels, with our value obtained from output stabilizers and theirs from the target codeword's Gaussian peak width. The $6.4$, $8.6$, and $10.3\,\mathrm{dB}$ examples in Table~I of Ref.~\cite{takase_gottesman-kitaev-preskill_2023} use PNR outcomes of $n=6$, $10$, and $16$, respectively. In the squeezed-single-photon route, increasing the resource size has a different cost. One requires more clocked seeds and additional homodyne-breeding layers, but no single detector is required to resolve a larger photon number.

Recently, iterative schemes have been proposed to enhance the success probability of non-Gaussian resource preparation \cite{winnel_deterministic_2024, erkilic_unified_2026}. However, they still require efficient PNR detectors and the generation rates are limited by the speed of optical switch.

The physical output rate depends on both the overall success probability and the achievable trial repetition rate. When high-$n$ photon-number resolution or optical switch limits a heralding implementation, efficient clocked single-photon sources and fast homodyne readout could make the present route competitive in generation rate. A quantitative rate comparison requires a common model of seed multiplexing, synchronization, and optical loss. The analysis here provides seed requirements and finite-window cat-preparation probabilities that can be incorporated into such a rate model.

\section{Conclusion}
\label{conclusion}
We have developed an analytic resource analysis of homodyne cat and GKP breeding starting from squeezed single-photon inputs. Standard homodyne breeding assumes non-Gaussian squeezed-cat states as input resources. Here we resolved this assumption one level further by asking whether the same Gaussian-and-homodyne breeding framework can be supported by squeezed single photons motivated by clocked single-photon sources. Without modifying the breeding circuit, we connect the single-photon seed, the bred cat resource, and the resulting GKP state within a single framework.

For the symmetric cat-breeding tree with $N=2^{m_1}$, the output reduces to a polynomial--Gaussian state controlled by one effective width. This reduction gives closed-form even-cat fidelities, an inverse-design rule for the seed squeezing, and the expected amplitude amplification across successive breeding rounds. Applying the resulting resources to GKP breeding shows that the one-step outputs can have large overlaps with finite-energy GKP states, while the stabilizer-based metrics reveal generally anisotropic effective squeezing in the two quadratures. Additional GKP-to-GKP breeding extends the grid structure and trades effective squeezing between the two quadratures at an increased seed cost. We also extended the analysis beyond the ideal conditioning point. For nonzero homodyne outcomes, the conditional state remains a finite polynomial--Gaussian object, allowing finite-window probabilities and averaged fidelities to be evaluated from Gaussian moments without sampling. This identifies an intermediate window where the heralding probability is finite and the cat fidelity remains useful. Photon loss analysis further shows that the spacing-matched resources relevant for GKP generation can be more stable than the bare cat fidelity alone suggests, although larger-amplitude resources are more fragile.

The resulting picture is a different assignment of the non-Gaussian resource cost within the established breeding framework. PNR-based cat and GKP generation schemes scale toward larger resources by demanding higher photon number resolution in the heralding detector. In contrast, the squeezed-single-photon implementation studied here avoids high-$n$ PNR measurements in the breeding stage and instead pays the cost through the preparation, synchronization, and Gaussian processing of many clocked seeds. This assignment is motivated primarily by recent progress in deterministic quantum-dot single-photon platforms, including both high efficiency and indistinguishability sources, programmable photonic
processing, telecom-band operation, and GHz-clocked emission~\cite{ding_high-efficiency_2025,maring_versatile_2024,hauser_deterministic_2026,rickert_high_2025}. Quadrature and intensity squeezing have also been observed in quantum dot resonance fluorescence~\cite{schulte_quadrature_2015,wang_observation_2020}. These developments make squeezed single-photon inputs a well motivated resource candidate for non-Gaussian state generation.

Several directions are left open. First, the achieved GKP squeezing falls slightly short of the best fault-tolerant threshold of 
$9.75\,\mathrm{dB}$ \cite{aghaee_rad_scaling_2025}. Although increasing the number of seed states could overcome this limitation, it would come at the cost of a reduced success probability. Since each additional layer in the symmetric breeding protocol doubles the number of seed states, developing a breeding protocol that accommodates $ N \neq 2^{m_1} $ would be desirable. On the other hand, further improvements in architecture design, including the choice of outer error-correcting code and decoding strategies, could lower the required GKP squeezing threshold. Next, a full rate model should combine the finite-window probabilities derived here with source multiplexing, feedforward, detector bandwidth, and a complete optical-loss budget. In this sense, the present work provides a resource level bridge between deterministic single-photon hardware and optical cat and GKP state generation.

\section*{acknowledgments}
This work is supported by the Korea Institute of Science and Technology (26E0001) and the National Research Foundation (NRF) of Korea (RS-2024-00509800).

\bibliography{references}

\clearpage
\onecolumngrid
\appendix

\section{Derivation of the cat breeding hierarchy and even-cat fidelity \label{AppendixA}}
\subsection{Two-seed conditional output \label{AppendixA1}}
We derive the conditional output of a single two-seed breeding step directly in the $\hat{x}$-quadrature wavefunction representation. The two squeezed single-photon seeds enter the beam splitter as the product wavefunction, 
\begin{equation}
\begin{aligned}
    \Psi_{\mathrm{in}}(x_1,x_2)&=\psi_{r_1}(x_1)\psi_{r_2}(x_2)\\ &\propto x_1x_2\exp\left[-\frac{1}{2}(e^{2r_1}x_1^2+e^{2r_2}x_2^2)\right].
\end{aligned}
\end{equation}
Writing $s=\sin\theta$ and $c=\cos\theta$, the beam-splitter transformation in the quadrature representation is given by
\begin{equation}
\begin{pmatrix} \hat{y}_1 \\ \hat{y}_2 \end{pmatrix} = 
\begin{pmatrix} c & s \\ -s & c \end{pmatrix} 
\begin{pmatrix} \hat{x}_1 \\ \hat{x}_2\end{pmatrix},
\end{equation}
with the same transformation applied to the $\hat{p}$ quadratures. The corresponding inverse transformation is $x_1=cy_1-sy_2,\quad x_2=sy_1+cy_2$. Substituting these relations rewrites the prefactor and Gaussian exponent in the output coordinates as 
\begin{equation}
\begin{aligned}
    x_1x_2&=cs(y^2_1-y^2_2)+(c^2-s^2)y_1y_2, \\[4pt]
    e^{2r_1}x^2_1+e^{2r_2}x^2_2&=Ay^2_1+By^2_2+2Cy_1y_2, \\[4pt]
    A=e^{2r_1}c^2+e^{2r_2}s^2,\quad B&=e^{2r_1}s^2+e^{2r_2}c^2,\quad C=cs(e^{2r_2}-e^{2r_1}).
\end{aligned}
\end{equation}
Hence, the beam-splitter output wavefunction is
\begin{equation}
    \Psi_{\mathrm{in}}(cy_1-sy_2,sy_1+cy_2)\propto \left[cs(y^2_1-y^2_2)+(c^2-s^2)y_1y_2\right] \exp\left[-\frac{1}{2}(Ay^2_1+By^2_2+2Cy_1y_2)\right].
\end{equation}
Conditioning mode $2$ on the homodyne outcome $p=0$ projects onto $\ket{p=0}$, and since $\braket{p=0|y_2}=1/\sqrt{2\pi}$ is independent of $y_2$, the projection reduces to an integration of $y_2$. This integral is Gaussian; evaluating it together with the first two $y_2$ moments gives
\begin{equation}
\begin{aligned}
    \int^\infty_{-\infty}dy_2\,e^{-\frac{1}{2}(By^2_2+2Cy_1y_2)}&=\sqrt{\frac{2\pi}{B}}e^{C^2y^2_1/2B},\quad \braket{y_2}=-\frac{Cy_1}{B},\quad \braket{y^2_2}=\frac{1}{B}+\frac{C^2y^2_1}{B^2},\\[4pt]
    \int^\infty_{-\infty}dy_2\,\Psi_{\mathrm{in}}(cy_1-sy_2,sy_1+cy_2)&\propto \left[csy^2_1-\frac{C}{B}\left(c^2-s^2\right)y^2_1-cs\left(\frac{1}{B}+\frac{C^2y^2_1}{B^2}\right)\right] \exp\left[-\frac{1}{2}\left(A-\frac{C^2}{B}\right)y^2_1\right], \\
    &\propto \left(1-\sigma_{\mathrm{eff}}y^2_1\right) \exp\left(-\frac{1}{2}\sigma_{\mathrm{eff}}y^2_1\right).
\end{aligned}
\end{equation}
Here the constant prefactor $-cs/B$ is absorbed into the normalization. The effective width is defined as
\begin{equation}
    \sigma_{\mathrm{eff}}=A-\frac{C^2}{B}=\frac{e^{2(r_1+r_2)}}{e^{2r_1}\sin^2\theta+e^{2r_2}\cos^2\theta}.
\end{equation}
The final $\pi/2$ phase-space rotation is equivalent to a Fourier transform. Applying it to the conditional state gives
\begin{equation}
    \psi^{(2)}_\mathrm{out}(x)\propto \frac{1}{\sqrt{2\pi}}\int_{-\infty}^\infty dy_1\,e^{-ixy_1}\left(1-\sigma_{\mathrm{eff}}y^2_1\right) \exp\left(-\frac{1}{2}\sigma_{\mathrm{eff}}y^2_1\right) \propto x^2\exp\left(-\frac{x^2}{2\sigma_{\mathrm{eff}}}\right).
\end{equation}
The normalization factor follows from
\begin{equation}
    \int_{-\infty}^\infty dx\,x^4\exp\left(-\frac{x^2}{\sigma_\mathrm{eff}}\right)=\frac{3\sqrt{\pi}}{4}\sigma_\mathrm{eff}^{5/2}.
\end{equation}
Thus, the normalized output wavefunction is
\begin{equation}
    \psi^{(2)}_\mathrm{out}(x)=\frac{2}{\sqrt{3}\pi^{1/4}\sigma_\mathrm{eff}^{5/4}} x^2\exp\left(-\frac{x^2}{2\sigma_\mathrm{eff}}\right),
\end{equation}
which reproduces Eq.~\eqref{N2_output}.

\subsection{Hermite recursion in $N=2^{m_1}$ cat breeding \label{AppendixA2}}
The two-seed state obtained in Appendix~\ref{AppendixA1} provides the initial state for the recursion. Before the final rotation, it can be written in the Hermite basis as
\begin{equation}
    \left(1-\sigma_{\mathrm{eff}}y^2\right) \exp\left(-\frac{1}{2}\sigma_{\mathrm{eff}}y^2\right)\propto H_2\left(y\sqrt{\frac{\sigma_\mathrm{eff}}{2}}\right) \exp\left(-\frac{1}{2}\sigma_{\mathrm{eff}}y^2\right).
\end{equation}
Here, $H_n(z)$ denotes the Hermite polynomial, 
\begin{equation}
    H_n(z)=(-1)^n e^{z^2}\frac{d^n}{dz^n}e^{-z^2}.
\end{equation}
Suppose that at some intermediate round the two input modes carry identical order-$n$ states,
\begin{equation}
    \phi_n(x)\propto H_n\left(x\sqrt{\frac{\sigma_{\mathrm{eff}}}{2}}\right)\exp\left(-\frac{1}{2}\sigma_{\mathrm{eff}}x^2\right).
\end{equation}
At the balanced beam splitter $\theta=\pi/4$, the point transformation becomes $x_1=(y_1-y_2)/\sqrt{2},\quad x_2=(y_1+y_2)/\sqrt{2}$. The relation $x^2_1+x^2_2=y^2_1+y^2_2$ leaves the Gaussian envelope untouched and mixes only the two Hermite factors. This gives
\begin{equation}
    \Phi_n(y_1,y_2)\propto H_n\left(\frac{y_1-y_2}{\sqrt{2}}\sqrt{\frac{\sigma_{\mathrm{eff}}}{2}}\right) H_n\left(\frac{y_1+y_2}{\sqrt{2}}\sqrt{\frac{\sigma_{\mathrm{eff}}}{2}}\right) \exp\left[-\frac{1}{2}\sigma_{\mathrm{eff}}(y^2_1+y^2_2)\right].
\end{equation}
Projecting mode $2$ onto $p=0$ once more amounts to an integration over $y_2$. We pass to the rescaled variables $X=y_1\sqrt{\sigma_{\mathrm{eff}}/2},\quad u=y_2\sqrt{\sigma_{\mathrm{eff}}/2}$, in which the two inputs read $H_n\left(\frac{X-u}{\sqrt{2}}\right)$ and $H_n\left(\frac{X+u}{\sqrt{2}}\right)$: the beam splitter splits $X$ between the two factors, with $u$ the integrated variable. The envelope likewise splits as $e^{-\sigma_{\mathrm{eff}}(y^2_1+y^2_2)/2}=e^{-X^2}e^{-u^2}$, so that $e^{-X^2}$ leaves the integral and only a Hermite product remains
\begin{equation}
    \int^\infty_{-\infty}dy_2\,\Phi_n(y_1,y_2)\propto e^{-X^2}\int^\infty_{-\infty}du\,e^{-u^2}H_n\left(\frac{X-u}{\sqrt{2}}\right)H_n\left(\frac{X+u}{\sqrt{2}}\right),
\end{equation}
whose closed form is supplied by the identity
\begin{equation}
    \int^{\infty}_{-\infty}du\,e^{-u^2} H_n\left(\tfrac{X-u}{\sqrt{2}}\right) H_n\left(\tfrac{X+u}{\sqrt{2}}\right) = \frac{\sqrt{\pi}}{2^n}H_{2n}(X).
\label{Hermite_product}
\end{equation}
We restore the $e^{-X^2}$ factor; the heralded mode is again a single Hermite-Gaussian, now raised to order $2n$ at the same width,
\begin{equation}
    \phi_{2n}(x)\propto H_{2n}(X)e^{-X^2} = H_{2n}\left(x\sqrt{\frac{\sigma_{\mathrm{eff}}}{2}}\right) \exp\left(-\frac{1}{2}\sigma_{\mathrm{eff}}x^2\right).
\end{equation}
The round therefore enacts $n\to2n$ with $\sigma_{\mathrm{eff}}$ held fixed. Starting from the two-seed state with $n=2$, each additional breeding round doubles the Hermite order. After $m_1-1$ additional rounds, the total number of seeds is $N=2^{m_1}$, and the pre-rotation state is given by
\begin{equation}
    \Phi^{(N)}_{\mathrm{pre}}(y)\propto H_N\left(y\sqrt{\frac{\sigma_{\mathrm{eff}}}{2}}\right)\exp\left(-\frac{1}{2}\sigma_{\mathrm{eff}}y^2\right).
\label{N_pre}
\end{equation}

The closing $\pi/2$ rotation realizes a Fourier transform of this state. Setting $z=y\sqrt{\sigma_{\mathrm{eff}}/2}$, the Hermite-Gaussian factor admits the Rodrigues form
\begin{equation}
    H_N(z)e^{-z^2}=(-1)^N\frac{d^N}{dz^N}e^{-z^2}.
\end{equation}
Because $z$ and $y$ differ only by the constant $\sqrt{\sigma_{\mathrm{eff}}/2}$, each $z$-derivative carries a fixed factor, and up to normalization the pre-rotation state is proportional to $\frac{d^N}{dy^N}e^{-\sigma_{\mathrm{eff}}y^2/2}$. Feeding this into the Fourier integral and integrating by parts $N$ times--the boundary terms vanishing through the Gaussian decay--transfers every derivative onto the kernel $e^{-ixy}$, each contributing a factor $ix$, 
\begin{equation}
    \int^\infty_{-\infty}dy\,e^{-ixy}\Phi^{(N)}_{\mathrm{pre}}(y)\propto \int^\infty_{-\infty}dy\,e^{-ixy}\frac{d^N}{dy^N}e^{-\sigma_{\mathrm{eff}}y^2/2}=(ix)^N\int^\infty_{-\infty}dy\,e^{-ixy}e^{-\sigma_{\mathrm{eff}}y^2/2}.
\end{equation}
The remaining factor is a Gaussian transform,
\begin{equation}
    \frac{1}{\sqrt{2\pi}}\int^\infty_{-\infty}dy\,e^{-ixy}e^{-\sigma_{\mathrm{eff}}y^2/2}=\frac{1}{\sqrt{\sigma_{\mathrm{eff}}}}\exp\left(-\frac{x^2}{2\sigma_{\mathrm{eff}}}\right).
\end{equation}
Gathering these factors recovers the monomial-Gaussian state of the main text,
\begin{equation}
    \psi^{(N)}_{\mathrm{out}}(x)=\mathcal{N}^{(N)}_{\sigma_{\mathrm{eff}}}x^N\exp\left(-\frac{x^2}{2\sigma_{\mathrm{eff}}}\right),
\end{equation}
where the prefactor is calculated by $\int dx|\psi^{(N)}_{\mathrm{out}}(x)|^2=1$,
\begin{equation}
    \mathcal{N}^{(N)}_{\sigma_{\mathrm{eff}}}=\frac{2^{N/2}}{\sqrt{(2N-1)!!}\pi^{1/4}\sigma_{\mathrm{eff}}^{(2N+1)/4}}.
\end{equation}

\subsection{Even-cat fidelity \label{AppendixA3}}
The even-cat fidelity follows from the overlap with one coherent component. Its $\hat{x}$ wavefunction is
\begin{equation}
    \braket{x|\alpha}=\pi^{-1/4}\exp\left[-\frac{1}{2}(x-\sqrt{2}\alpha)^2\right].
\end{equation}
We then contract it against the heralded output to obtain a Gaussian-weighted moment integral,
\begin{equation}
    \braket{\alpha|\psi^{(N)}_{\mathrm{out}}}=\mathcal{N}^{(N)}_{\sigma_{\mathrm{eff}}}\pi^{-1/4}\int^\infty_{-\infty}dx\,x^N\exp\left[-\frac{1}{2}(x-\sqrt{2}\alpha)^2-\frac{x^2}{2\sigma_{\mathrm{eff}}}\right].
\end{equation}
Completing the square merges the two Gaussians into one of mean $\bar{x}$ and variance $V_x$, with a constant factor $e^{-\alpha^2/(\sigma_{\mathrm{eff}}+1)}$ left over,
\begin{equation}
\begin{aligned}
    -\frac{1}{2}(x-\sqrt{2}\alpha)^2-\frac{x^2}{2\sigma_{\mathrm{eff}}}&=-\frac{(x-\bar{x})^2}{2V_x}-\frac{\alpha^2}{\sigma_{\mathrm{eff}}+1}\\[4pt]
    \bar{x}=\frac{\sqrt{2}\alpha\sigma_{\mathrm{eff}}}{\sigma_{\mathrm{eff}}+1}&,\quad V_x=\frac{\sigma_{\mathrm{eff}}}{\sigma_{\mathrm{eff}}+1}.
\end{aligned}
\end{equation}
The remaining integral is the $N$th moment of this shifted Gaussian, a finite sum evaluated with $(-1)!!=1$,
\begin{equation}
    \int^\infty_{-\infty}dx\,x^N\exp\left[-\frac{(x-\bar{x})^2}{2V_x}\right]=\sqrt{2\pi V_x}\sum^{N/2}_{k=0}\binom{N}{2k}(2k-1)!!V^k_x\bar{x}^{N-2k}.
\end{equation}
We insert $\bar{x}$ and $V_x$ to collect the $\alpha$ and $\sigma_{\mathrm{eff}}$ dependence into a single polynomial,
\begin{equation}
    M_N(\sigma_{\mathrm{eff}},\alpha)=\sum^{N/2}_{k=0}\binom{N}{2k}(2k-1)!!(\sigma_{\mathrm{eff}}+1)^k(2\alpha^2\sigma_{\mathrm{eff}})^{\frac{N}{2}-k},
\end{equation}
so the squared single-component overlap becomes
\begin{equation}
    \left|\braket{\alpha|\psi^{(N)}_{\mathrm{out}}}\right|^2=2\sqrt{\pi}\left(\mathcal{N}^{(N)}_{\sigma_\mathrm{eff}}\right)^2\frac{\sigma_\mathrm{eff}^{N+1}}{(\sigma_\mathrm{eff}+1)^{2N+1}}[M_N(\sigma_\mathrm{eff},\alpha)]^2\exp\left(-\frac{2\alpha^2}{\sigma_\mathrm{eff}+1}\right).
\end{equation}
Since $\psi^{(N)}_{\mathrm{out}}$ has even parity, the $-\alpha$ component contributes identically, $\braket{\alpha|\psi^{(N)}_{\mathrm{out}}}=\braket{-\alpha|\psi^{(N)}_{\mathrm{out}}}$. The even-cat fidelity is the overlap with the normalized even cat,
\begin{equation}
\begin{aligned}
    \ket{\mathrm{Cat}_+(\alpha)}=\frac{1}{\mathcal{N}_+}(\ket{\alpha}+\ket{-\alpha})&,\quad \mathcal{N}_+=\sqrt{2[1+\exp(-2\alpha^2)]},\\[4pt]
    F^{(N)}_+(\sigma_{\mathrm{eff}},\alpha)&=\left|\braket{\mathrm{Cat}_+(\alpha)|\psi^{(N)}_{\mathrm{out}}}\right|^2.
\end{aligned}
\end{equation}
Including the normalization of the even cat state, we obtain the closed-form fidelity
\begin{equation}
    F^{(N)}_+(\sigma_{\mathrm{eff}},\alpha)=\frac{C_N\sqrt{\sigma_{\mathrm{eff}}}[M_N]^2}{(\sigma_{\mathrm{eff}}+1)^{2N+1}}\frac{e^{2\sigma_{\mathrm{eff}}\alpha^2/(\sigma_{\mathrm{eff}}+1)}}{e^{2\alpha^2}+1},\quad C_N=\frac{2^{N+2}}{(2N-1)!!}.
\label{N_Fidelity}
\end{equation}
For $N=2$, the polynomial and prefactor reduce to
\begin{equation}
    M_2(\sigma_\mathrm{eff},\alpha)=2\sigma_\mathrm{eff}\alpha^2+\sigma_\mathrm{eff}+1,\quad C_2=\frac{16}{3}.
\end{equation}
Substituting these expressions into Eq.~\eqref{N_Fidelity} reproduces the two-seed fidelity in Eq.~\eqref{N2_Fidelity}.

\section{Derivation of the finite-outcome state}
\label{AppendixB}
\subsection{Orthogonal network and the combined Fourier phase}
We work in the symmetric configuration, in which the $N$-mode input before the beam-splitter tree is
\begin{equation}
    \Psi_{\mathrm{in}}(\mathbf{x})\propto \left(\prod^N_{j=1}x_j\right)\exp\left[-\frac{\sigma_{\mathrm{eff}}}{2}\mathbf{x}^T\mathbf{x}\right],
\end{equation}
with $\mathbf{x}=(x_1,\dots,x_N)^T$ the input quadrature coordinates. All beam splitters are balanced, so the full network is a real orthogonal map $\mathbf{z}=O_N\mathbf{x}$, with $O^T_NO_N=I_N$. We order the output coordinates as $\mathbf{z}=(y,v_1,\dots,v_{N-1})^T$, where $y$ is the surviving mode and $v_1,\dots,v_{N-1}$ are the modes measured across the $m$ layers of the tree. Orthogonality preserves the quadratic form,
\begin{equation}
    \mathbf{x}^T\mathbf{x}=\mathbf{z}^T\mathbf{z}=y^2+\sum^{N-1}_{\ell=1}v^2_\ell,
\end{equation}
leaving the Gaussian envelope invariant and only the polynomial factor is transformed by the network. For a general outcome $p_\ell$, each measured mode is projected onto a $\hat{p}$-quadrature eigenstate, $\braket{p_\ell|v_\ell}\propto e^{-ip_\ell v_\ell}$, and the final $\pi/2$ rotation of the surviving mode is a Fourier transform with kernel $e^{-ixy}$. If we collect the conjugate variables into $\mathbf{q}=(x,p_1,\dots,p_{N-1})^T$, these phases combine as
\begin{equation}
    e^{-i(xy+\sum^{N-1}_{\ell=1} p_\ell v_\ell)}=e^{-i\mathbf{q}^T\mathbf{z}}=e^{-i\mathbf{q}^TO_N\mathbf{x}}=e^{-i(O^T_N\mathbf{q})^T\mathbf{x}}.
\end{equation}
The heralded state is thus the $N$-dimensional Gaussian Fourier transform of $\Psi_{\mathrm{in}}$ evaluated at the vector $\mathbf{k}=O^T_N\mathbf{q}$.

\subsection{Factorization of the Gaussian Fourier transform}
As the phase is diagonal in $\mathbf{x}$, the transform separates across the input coordinates,
\begin{equation}
    \tilde{\psi}_N(x;\mathbf{p})\propto\int^\infty_{-\infty}d^N\mathbf{x}\,\left(\prod^N_{j=1}x_j\right)e^{-\sigma_{\mathrm{eff}}\mathbf{x}^T\mathbf{x}/2}e^{-i\mathbf{k}^T\mathbf{x}}\propto\prod^N_{j=1}\int^\infty_{-\infty}dx_j\,x_je^{-\sigma_{\mathrm{eff}}x^2_j/2}e^{-ik_jx_j}.
\end{equation}
Each one-dimensional integral is a first-moment Gaussian transform,
\begin{equation}
    \int^\infty_{-\infty}dx_j\,x_je^{-\sigma_{\mathrm{eff}}x^2_j/2}e^{-ik_jx_j}\propto k_j\exp\left(\frac{-k^2_j}{2\sigma_{\mathrm{eff}}}\right).
\end{equation}
Taking the product over $j$ and discarding $\mathbf{q}$-independent constants gives
\begin{equation}
    \tilde{\psi}_N(x;\mathbf{p})\propto\left(\prod^N_{j=1}k_j\right)\exp\left(-\frac{1}{2\sigma_{\mathrm{eff}}}\sum^N_{j=1}k^2_j\right).
\end{equation}
The orthogonality of $O_N$ leaves the exponent unchanged,
\begin{equation}
    \sum^N_{j=1}k^2_j=\mathbf{k}^T\mathbf{k}=\mathbf{q}^T\mathbf{q}=x^2+|\mathbf{p}|^2,
\end{equation}
then the heralded state takes the closed form
\begin{equation}
    \tilde{\psi}_N(x;\mathbf{p})\propto Q_N(x;\mathbf{p})\exp(-\frac{x^2+|\mathbf{p}|^2}{2\sigma_{\mathrm{eff}}}),\quad Q_N(x;\mathbf{p})=N^{N/2}\prod^N_{j=1}(O^T_N\mathbf{q})_j.
\end{equation}
Here, the prefactor $N^{N/2}$ is fixed by the $\mathbf{p}=0$ limit:
$Q_N(x;\mathbf{0})=N^{N/2}(x/\sqrt{N})^N=x^N$.

\subsection{Success probability and average fidelity}
The continuous-variable homodyne measurement accepts any outcome within the box window $\mathcal{W}(w)=[-w,w]^{N-1}$. The probability density of an outcome $\mathbf{p}$ is the norm of the unnormalized conditional state,
\begin{equation}
    P_N(\mathbf{p})\propto\int^\infty_{-\infty}dx|\tilde{\psi}_N(x;\mathbf{p})|^2=e^{-|\mathbf{p}|^2/\sigma_{\mathrm{eff}}}\mathcal{N}_N(\mathbf{p}),\quad\mathcal{N}_N(\mathbf{p})=\int^\infty_{-\infty}dx|Q_N(x;\mathbf{p})|^2e^{-x^2/\sigma_{\mathrm{eff}}},
\end{equation}
so the normalized acceptance probability, the fraction of outcomes falling inside the window, is
\begin{equation}
    P^{(N)}_{\mathrm{acc}}(w)=\frac{\int_{\mathcal{W}(w)}d^{N-1}\mathbf{p}\,e^{-|\mathbf{p}|^2/\sigma_{\mathrm{eff}}}\mathcal{N}_N(\mathbf{p})}{\int_{\mathbb{R}^{N-1}}d^{N-1}\mathbf{p}\,e^{-|\mathbf{p}|^2/\sigma_{\mathrm{eff}}}\mathcal{N}_N(\mathbf{p})}.
\end{equation}
Both integrals are calculated analytically rather than by sampling. Since $Q_N(x;\mathbf{p})$ is a finite polynomial,
\begin{equation}
    Q_N(x;\mathbf{p})=\sum_\mathbf{a}c_\mathbf{a}x^{a_0}p^{a_1}_1p^{a_2}_2\dots p^{a_{N-1}}_{N-1},
\end{equation}
the squared polynomial $|Q_N|^2$ is also finite. The acceptance probability and the fidelity can then be computed from the one-dimensional Gaussian moments
\begin{equation}
    \int^\infty_{-\infty}dx\,x^{2k}e^{-x^2/\sigma_{\mathrm{eff}}},\quad\int^w_{-w}dp\,p^{2k}e^{-p^2/\sigma_{\mathrm{eff}}},\quad\int^\infty_{-\infty}dp\,p^{2k}e^{-p^2/\sigma_{\mathrm{eff}}},
\end{equation}
where the $x$-integral evaluates the overlaps, the windowed $p$-integral covers the acceptance window $\mathcal{W}(w)$, and the full-range $p$-integral normalizes over all outcomes. The same reduction applies to the fidelity: at each outcome the heralded state is overlapped with the target even cat to give the outcome-dependent fidelity $F_N(\mathbf{p})=|\braket{\mathrm{Cat}_+(\alpha)|\psi_N(\mathbf{p})}|^2$, and weighting by the outcome probability yields the window-averaged fidelity,
\begin{equation}
    \bar{F}_N(w)=\frac{\int_{\mathcal{W}(w)}d^{N-1}\mathbf{p}\,P_N(\mathbf{p})F_N(\mathbf{p})}{\int_{\mathcal{W}(w)}d^{N-1}\mathbf{p}\,P_N(\mathbf{p})}.
\end{equation}

\subsection{N=2,4,8 outcome polynomials}
We list the orthogonal network matrix and the resulting outcome polynomial for $N=2,4,8$. The single-layer tree $N=2$ is one balanced beam splitter, measuring the difference mode $p_1$:
\begin{equation}
    O_2=\frac{1}{\sqrt{2}}\begin{pmatrix}1&1\\1&-1\end{pmatrix},\quad\mathbf{q}=(x,p_1)^T,\quad Q_2(x;\mathbf{p})=(x+p_1)(x-p_1).
\end{equation}
The two-layer tree $N=4$ adds the differences $p_1, p_2$ measured at first layer and $p_3$ at the second:
\begin{equation}
    O_4=\frac{1}{2}\begin{pmatrix}1&1&1&1\\\sqrt{2}&-\sqrt{2}&0&0\\0&0&\sqrt{2}&-\sqrt{2}\\1&1&-1&-1\end{pmatrix},\quad\mathbf{q}=(x,p_1,p_2,p_3)^T,\quad Q_4(x;\mathbf{p})=\left[(x+p_3)^2-2p^2_1\right] \left[(x-p_3)^2-2p^2_2\right].
\end{equation}
The three-layer tree $N=8$ follows by the same construction,
\begin{equation}
    O_8=\frac{1}{2\sqrt{2}}\begin{pmatrix}1&1&1&1&1&1&1&1\\2&-2&0&0&0&0&0&0\\0&0&2&-2&0&0&0&0\\0&0&0&0&2&-2&0&0\\0&0&0&0&0&0&2&-2\\\sqrt{2}&\sqrt{2}&-\sqrt{2}&-\sqrt{2}&0&0&0&0\\0&0&0&0&\sqrt{2}&\sqrt{2}&-\sqrt{2}&-\sqrt{2}\\1&1&1&1&-1&-1&-1&-1\end{pmatrix},\quad\mathbf{q}=(x,p_1,\dots,p_7)^T
\end{equation}
with outcome polynomial
\begin{equation}
\begin{aligned}
    Q_8(x;\mathbf{p}) = & \left[(x+\sqrt{2}p_5+p_7)^2-4p_1^2\right] \left[(x-\sqrt{2}p_5+p_7)^2-4p_2^2\right]\\
    & \times \left[(x+\sqrt{2}p_6-p_7)^2-4p_3^2\right] \left[(x-\sqrt{2}p_6-p_7)^2-4p_4^2\right].
\end{aligned}
\end{equation}
The pattern continues recursively for larger $N$; the $N=16$ polynomial is generated in the same way and used in the results of Sec.~\ref{finite_window}.

\end{document}